\documentclass{aa}
\usepackage[english]{babel}
\usepackage[varg]{txfonts}

\usepackage[usenames,dvipsnames]{xcolor}

\usepackage{epsfig}
\usepackage{graphicx,natbib}
\usepackage{amsmath,amssymb}
\usepackage{hyperref}
\usepackage{wasysym}
\usepackage{natbib}
\usepackage{ulem}
\usepackage{cancel}

\hypersetup{
    colorlinks=true,
    citecolor=blue,
    linkcolor=red,
    urlcolor=black
    }  

\def\ga{\,\hbox{\hbox{$ > $}\kern -0.8em \lower 1.0ex\hbox{$\sim$}}\,}
\def\la{\,\hbox{\hbox{$ < $}\kern -0.8em \lower 1.0ex\hbox{$\sim$}}\,}
\def\aap{A\&A}

\newcommand{\vect}[1]{\mathbf{#1}}

\newcommand{\demi}{\frac{1}{2}}

\newcommand{\trm}[1]{\mathrm{#1}}

\newcommand{\refig}[1]{Fig.~\ref{#1}}
\newcommand{\refigs}[2]{Figs.~\ref{#1} and \ref{#2}}

\newcommand{\cc}{{\rm cm}^{-3}}

\newcommand{\noprint}[1]{{}}

\begin{document}
\def\nat{Nature }
\def\apj{Astrophys. J. }
\def\apjs{Astrophys. J., Suppl. Ser. }
\def\apjl{Astrophys. J., Lett. }
\def\apss{Astrophys. and Space Science}

\title{Chemical solver to compute molecule and grain abundances and non-ideal MHD resistivities in prestellar core-collapse calculations}
\titlerunning{A chemical solver to compute non-ideal MHD resistivities}

\author{P. Marchand \inst{1} 
       \and J. Masson \inst{1}
       \and G. Chabrier \inst{1,2}
       \and P. Hennebelle \inst{3}   
       \and B. Commer\c con\inst{1}
       \and N. Vaytet\inst{1,4}}

\institute{Ecole normale sup\'erieure de Lyon, CRAL, UMR CNRS 5574, 69364 Lyon Cedex 07,  France
\and School of Physics, University of Exeter, Exeter, EX4 4QL, UK
\and Laboratoire AIM, Paris-Saclay, CEA/IRFU/SAp-CNRS-Université Paris Diderot, 91191 Gif-sur-Yvette Cedex, France
\and Centre for Star and Planet Formation, Niels Bohr Institute and Natural History Museum of Denmark, University of Copenhagen, {\O}ster Voldgade 5-7, DK-1350 Copenhagen K, Denmark
}

\authorrunning{P. Marchand et~al.}

\date{}

\abstract{
We develop a detailed chemical network relevant to calculate the conditions that are characteristic of prestellar core collapse. We solve the system of time-dependent 
differential equations to calculate the equilibrium abundances of molecules and dust grains, 
with a size distribution given by size-bins for these latter. These abundances are used to compute the different non-ideal magneto-hydrodynamics resistivities (ambipolar, Ohmic and Hall), needed to carry out simulations of protostellar collapse. For the first time in this context, we take into account the evaporation of the grains,
the thermal ionisation of Potassium, Sodium and Hydrogen at high temperature, and the thermionic emission of grains
 in the chemical network, and we explore the impact of various cosmic ray ionisation rates. All these processes significantly affect the non-ideal magneto-hydrodynamics resistivities, which will modify the dynamics of the collapse. 
Ambipolar diffusion and Hall effect dominate at low densities, up to $n_\mathrm{H} = 10^{12}$ cm$^{-3}$, after which Ohmic diffusion takes over. We find that the time-scale needed to reach chemical equilibrium is always shorter than the typical dynamical (free fall) one. This allows us to build a large, multi-dimensional multi-species equilibrium abundance table over a large temperature, density and ionisation rate ranges. This table, which we make accessible to the community, is used during first and second prestellar core collapse calculations to compute the non-ideal magneto-hydrodynamics resistivities, yielding a consistent dynamical-chemical description of this process.\protect\thanks{This table is available in electronic form at the CDS via anonymous ftp to cdsarc.u-strasbg.fr (130.79.128.5) or via http://cdsweb.u-strasbg.fr/cgi-bin/qcat?J/A+A/}}

\maketitle

\section{Introduction}

It is well established today that stars form within turbulent complexes that range from tens to hundreds
of parsecs, that harbour thousands of solar masses of cold gas: the molecular clouds \protect\citep[e.g.][ for a review]{McKeeOstriker2007}. Turbulence within these clouds generates
over-dense regions where the balance between the thermal and non-thermal (outward) pressure and the gravitational
(inward) force is disrupted, causing the gas to collapse. The mass distribution of these dense cores
shows many similarities with the stellar mass function \citep[see e.g.][]{Konyves2010}, making them the most obvious stellar progenitors. The details of the formation process are still actively debated however. The
(initially optically thin) core is believed to first contract isothermally as the compression heating is lost
through radiation, until the density is high enough to render the cooling ineffective. This leads to the
formation of a hydrostatic body, known as the first Larson core \protect\citep{Larson1969}, which accretes material from the surrounding envelope.
 The sustained increase in mass, density, and temperature eventually triggers the dissociation of $\mathrm{H}_{2}$ molecules above 2000 K. This leads to the second phase of collapse
because of the endothermic nature of the dissociation process.
The collapse ceases when most or all of the $\mathrm{H}_{2}$ molecules
have been dissociated, at which point a second much more dense and compact hydrostatic core (Larson's second core) is formed at the centre \protect\citep{Larson1969,Masunaga2000,2013arXiv1307.1010V}.
The temperature inside the second core continues to rise until the nuclear reactions are ignited: the young
star is born.

This problem entails many physical processes over a very wide range of spatial scales, making
any modelling of the entire process challenging; the cloud core has an initial radius of $\sim$10$^{4}$ astronomical
units (AU)
with an average density of a few $10^3~\cc$, while the protostellar core measures only $10^{-3}$ AU at birth,
with densities over $10^{20}~\cc$. Describing and understanding
the physical processes at work involves an intricate interplay between large-scale environmental factors,
which regulate the supply of mass, angular momentum, and magnetic flux, and small-scale processes, which control the evolution and dynamics in protostellar systems. An accurate description of a large number of complex
physical mechanisms, involving the magnetic field, gravity, radiative transfer, time-dependent chemistry,
and dust physics, is necessary to derive realistic models of the global star formation process.

Recently, various studies have been devoted to the role magnetic fields play in collapsing systems and the effect on the transport of angular momentum. 
The first studies were based on ideal magnetohydrodynamics (MHD), implying numerical instead of physical diffusivity. The results concerning the disk formation are thus of dubious validity \citep{AllenShuLi,2004ApJ...616..266M,GalliShuLizano2006,PriceBate2007,HennebelleTeyssier2008,HennebelleFromang2008,commercon10,DA1}.
 The reconnection and diffusion of the field are essential components of the magnetic braking process. Using the framework of non-ideal MHD allows describing the diffusivity and the magnetic properties of the charged fluid accurately\citep{Machida_etal06,DuffinPudritz,MellonLi2009,MachidaMatsumotoDisk}.
This requires the accurate calculations of magnetic resistivities, however. 
This remains challenging as they depend on many factors such as temperature, density, chemical abundances and magnetisation.

The interstellar gas from which the cores (and thus stars) form is essentially composed of hydrogen ($\sim$74\% by mass), helium ($\sim$25\%), and 
heavier elements ($\sim$1\%) such as carbon, oxygen and heavy metals (Na, Fe, etc.), whose respective abundances are determined by an elaborate set of chemical reactions. The complexity is even increased by of the
formation of dust grains, which arise from the aggregation of several molecules and can reach micrometer sizes \citep{KunzMouschovias2009}. Not only do grains react with other elements, they also carry electric charges 
and play the role of catalysts for other chemical reactions. 
The grain ionisation rates, or even the chemical reaction rates themselves, remain poorly constrained in environments typical of prestellar core collapse. 
The relative abundances and the degree of ionisation of the grains and molecules define the magnetic resistivities that regulate the dissipation of the magnetic flux through ambipolar and Ohmic diffusion and the Hall effect \citep{Krasnopolsky,Machida_etal06,LiKrasnopolskyShang}. 
These dissipation processes have a fundamental role in the formation and evolution of the first and second Larson cores and their associated disks and outflows. In particular, the interplay between flux-freezing and condensation of the global angular momentum strongly depends on whether ambipolar diffusion, Hall effect, or Ohmic diffusion is the dominant process.
To compute the accurate resisitivities under the conditions typical of collapsing molecular clouds, we have developed 
 a relevant reduced chemical network. This allows us to test the effect of
various parameters, such as the equilibrium abundances of chemical species, the evolution of grains of different sizes, and the cosmic-ray ionisation rate on the various non-ideal MHD diffusion coefficients. 
This network significantly extends these and other previous prescriptions \citep[]{UmebayashiNakano1990,NishiNakanoUmebayashi1991,Wardle1999,nakano2002,Wardle2007,KunzMouschovias2009,IlgnerNelson,Bai2011} 
by including new pieces of physics that are necessary to precisely describe the chemical evolution of the interstellar gas. 

The paper is organised as follows. In Sect. 2, we outline the main processes and physical ingredients relevant to star formation. In Sect. 3, we present our chemical network and the numerical method used to solve the reaction equations.
The resulting resistivities, the effect of each ingredient on the chemical equilibrium of the gas, and the consequences for star-forming systems are discussed in Sect. 4, while Sect. 5 is devoted to the conclusion.

\section{Physics of collapsing prestellar cores}

\subsection{Magnetic resistivities}\label{resistivities}

The complete induction equation in non-ideal MHD reads
\begin{align}
\frac{\partial \mathbf{B}}{\partial t} - \nabla \times ( \mathbf{v} \times \mathbf{B} )  = 
       -\frac{c^2}{4\pi}& \nabla \times \Bigg[  \eta_{\Omega} (\nabla \times \mathbf{B}) \nonumber \\
      +& \eta_\mathrm{H} \left\{ (\nabla \times \mathbf{B})\times \frac{\mathbf{B}}{||\vect{B}||} \right\} \nonumber \\
  +& \eta_\mathrm{AD} \frac{\mathbf{B}}{||\vect{B}||}  \times \left\{ (\nabla \times \mathbf{B})\times \frac{\mathbf{B}}{||\mathbf{B}||} \right\}  \Bigg] \label{induc_1},
\end{align}
where $\mathbf{B}$ denotes the magnetic field, $||\vect{B}||$ its $\mathrm{L}_2$ norm, and $\mathbf{v}$ the fluid velocity. 
$\eta_{\Omega}$, $\eta_{\mathrm{H}}$ and $\eta_{\mathrm{AD}}$ represent the Ohmic, Hall, and ambipolar resistivities, respectively. These dissipative terms account for collisions between neutral and charged species. They strongly depend on the chemical equilibrium and thus on the thermodynamic conditions. In contrast, ideal MHD considers the fluid as a mixture of perfectly coupled charged fluids, which corresponds to $\eta_{\Omega} = \eta_{\mathrm{H}} = \eta_{\mathrm{AD}} = 0$, ensuring flux-freezing of the magnetic field. 

The resistivities are defined in terms of the conductivities of the gas-dust mixture as
\begin{align}
\eta_{\Omega} &= \frac{1}{\sigma_{\parallel}}, \\
\eta_\mathrm{H} &= \frac{\sigma_\mathrm{H}}{\sigma_{\bot}^2+ \sigma_\mathrm{H}^2}, \\
\eta_\mathrm{AD} &= \frac{\sigma_{\bot}}{\sigma_{\bot}^2+ \sigma_\mathrm{H}^2} - \frac{1}{\sigma_{\parallel}},
\end{align}
where the
parallel, perpendicular, and Hall conductivites are in turn expressed, respectively, as
\begin{align}
\sigma_{\parallel} &= \sum_i \sigma_i, \\
\sigma_{\bot} &= \sum_i \frac{\sigma_i}{1+(\omega_i \tau_{i\mathrm{n}})^2}, \\
\sigma_\mathrm{H} &= -\sum_i \frac{\sigma_i \omega_i \tau_{i\mathrm{n}}}{1+(\omega_i \tau_{i\mathrm{n}})^2}.
\end{align}
with
$\displaystyle \sigma_i = \frac{n_i q_i^2 \tau_{i\mathrm{n}}}{m_i}$, $\displaystyle \omega_i = \frac{q_i B}{m_i c}$ the cyclotron frequency, and
\begin{equation}
\tau_{i\mathrm{n}} = \frac{1}{a_{i\mathrm{He}}} \frac{m_i + m_\mathrm{H_2}}{m_\mathrm{H_2}} \frac{1}{n_\mathrm{H_2} \langle \sigma_\mathrm{coll} w \rangle_{i}}. \label{defpoursigma}
\end{equation}
Here $i$ stands for any charged particle of charge $q_i$, with a charge particle number-density $n_i$, $m$ denotes the mass of a particle, 
and $c$ is the speed of light. The factor $a_{i\mathrm{He}}$ accounts for collisions with helium atoms and is equal to 1.14 for ions, 1.16 for electrons and 1.28 for grains \citep{DeschMouschovias}.
$\langle \sigma_\mathrm{coll} w \rangle_{i}$ is the rate constant for collisions between a particle $i$ and H$_2$ molecules\citep{DeschMouschovias,PintoGalli2}. 
The values are taken from \citet{PintoGalli2}. 
We note that these rates were calculated in a three-fluid formalism, and we use them in a multifluid context, but their dependance with temperature is a better approximation than the Langevin model \citet{PintoGalli2}.
$\omega_i$ can be negative, depending on the charge $q_i$. This is important when calculating the conductivities, as the Hall conductivity $\sigma_\mathrm{H}$ 
can be either positive or negative.
This implies that the Hall resistivity can be negative, since it is of the same sign as $\sigma_\mathrm{H}$.

  \subsection{Density and temperature range}

\begin{figure}
\begin{center}
\includegraphics[trim = 2cm 2cm 2cm 2cm,width=0.5\textwidth]{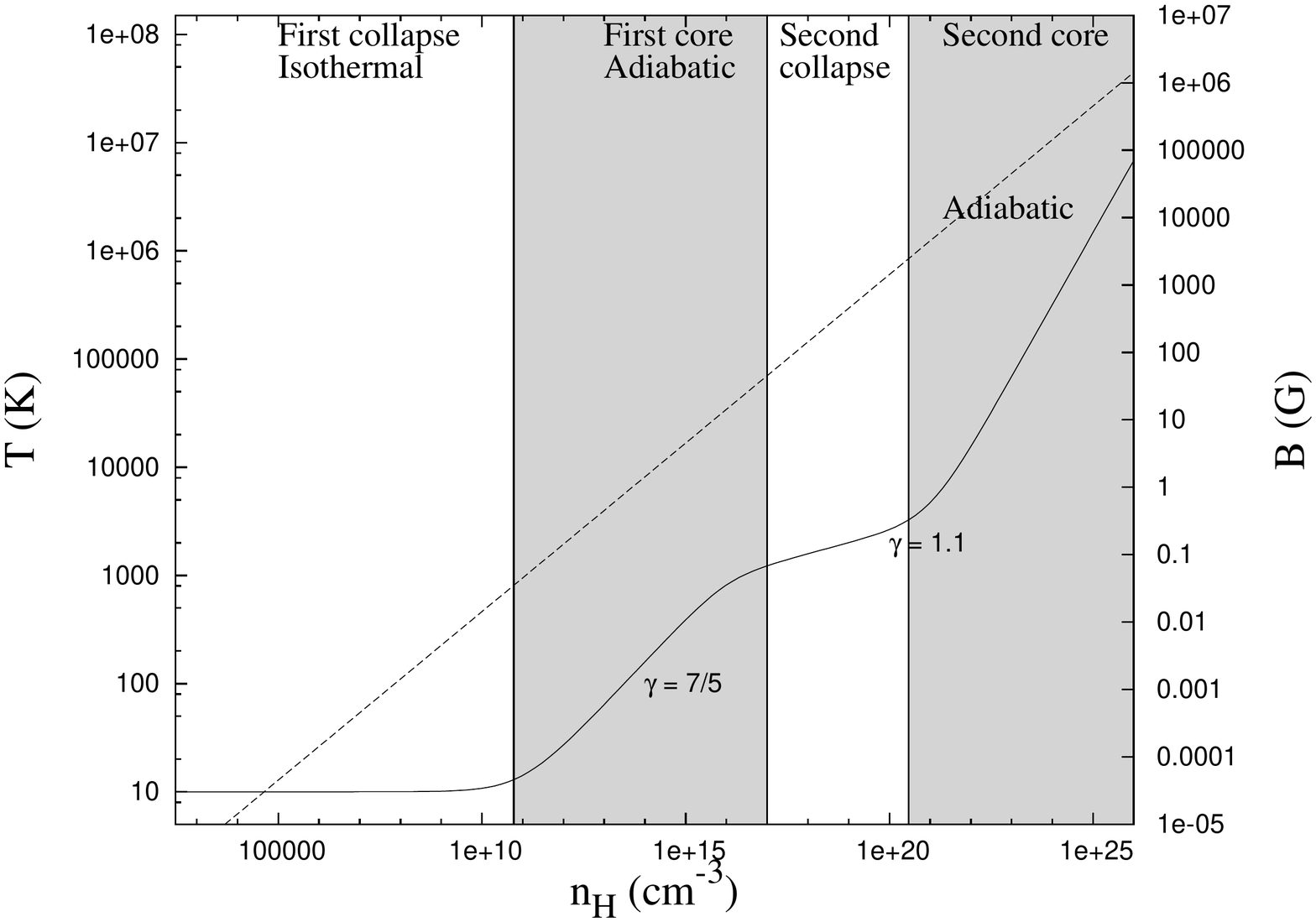}
\caption{Solid line : Temperature as a function of density for the equation of state. Dashed line : Magnetic field prescription following \citet{LiKrasnopolskyShang}}
\label{eosmag}
\end{center}
\end{figure}

Our chemical network was used to compute conductivities for an entire density-temperature range spanning $10^{2}~\text{cm}^{-3} < n < 10^{26}~\text{cm}^{-3}$ and $10~\text{K} < T < 10^{5}~\text{K}$. This covers typical (parent) molecular cloud conditions (low temperature and density) as well as the interior of first and second Larson cores, where densities and temperatures have increased by many orders of magnitude, as a result of the strong gravitational compression of the gas. However, for clarity as well as for comparison with previous studies, the majority of the results presented in this work are presented only as a function of density. To mimic the natural temperature evolution as a function of density in a collapsing dense core (as opposed to using a single constant temperature), we used the piecewise barotropic equation of state (EOS) from \citet{Machida_etal06}

\begin{equation}
T=T_0 \sqrt{1+\left(\frac{n}{n_1}\right)^{2 g_1}} \left(1+\left(\frac{n}{n_2}\right)\right)^{g_2} \left(1+\left(\frac{n}{n_3}\right)\right)^{g_3},
\end{equation}
with $n$ the total density and

\begin{equation}
\begin{array}{l@{~}c@{~}l@{~}c@{~}l@{~}c@{~}l}
n_1 &=& 10^{11}~\cc &; ~~~~~~~~~~~ & g_1 &=& 0.4; \\
n_2 &=& 10^{16}~\cc &; ~~~~~~~~~~~ & g_2 &=& -0.3; \\
n_3 &=& 10^{21}~\cc &; ~~~~~~~~~~~ & g_3 &=& 0.56667,
\end{array}
\end{equation}
and $T_0=10$ K. This EOS successively characterises the isothermal phase of the collapse, the adiabatic phase during the first Larson core evolution, the second collapse and the second Larson core evolution.
It is important to use such a typical EOS to present our results because the temperature increase during the collapse strongly influences on the chemistry of the grains 
(see Sect. \ref{grains_vap_sect}) and on the ionisation of potassium (Sect. \ref{network_sect}). 
The EOS is represented in \refig{eosmag} with our magnetic field prescription (see next paragraph).

\subsection{Magnetic field} \label{magfield}

Computing the resistivities also requires a knowledge of the magnetic field intensity.
To present our magnetic resistivities as a function of density alone, we assumed that the magnetic intensity scales as $B(n_\mathrm{H}) = 1.43 \times 10^{-7} \sqrt{n_\mathrm{H}}$ \citep{LiKrasnopolskyShang}, which corresponds to magnetic flux conservation (flux-freezing approximation). 
It is represented in \refig{eosmag}.
Although the scaling in real collapse calculations is more complicated \citep[see e.g.][]{DA1}, the magnetic field here is only used to illustrate the behaviour of the resistivities.
It has no influence on the chemistry. In detailed prestellar collapse calculations, the magnetic field evolution is properly calculated and the resistivities are computed consistently during the collapse \citep[see][]{DA1}.

\subsection{Dust grain model}\label{grains_sect}

Grains can be the main charge carriers and thus need to be accurately described, both in size and number density, because these two quantities determine the surface area available for chemical reactions. Furthermore, grain evaporation is a process of prime importance in the present context because it occurs at temperatures close to the second collapse, after the first core formation.

\subsubsection{Grain size}\label{grains_size}

The grain reaction rates and the conductivity of the dust-gas mixture depend on the grain cross section.
In our calculations, we included a power-law grain-size distribution by considering a finite number $N_{\mathrm{bins}}$ of size bins with equal widths in log space.
We define a minimum and maximum grain size $a_{\mathrm{min}}$ and $a_{\mathrm{max}}$, respectively, 
with a number density of grains of radius between $a$ and $a+da$

\begin{equation}
    dn_{\mathrm{g,tot}}(a)= \mathcal{C} \, a^{\lambda} \,da~,
\end{equation}
where the subscript g denotes the grains and $\mathcal{C}$ is a normalisation constant.
Unless otherwise stated, we used the standard MRN distribution with $\lambda = -3.5$ \citep{mathis} throughout this work. For the sake of generality,
 however, we write here the equations for any power-law index $\lambda$.
Following \citet{KunzMouschovias2009}, we chose for the minimum and maximum radii $a_{\mathrm{min}} = 0.0181 \, \mu\trm{m}$ and $ a_{\mathrm{max}} = 0.9049  \, \mu\trm{m}$.
Each size bin is defined by a lower and upper radius that give the number density and size for the $\alpha$th bin ($\alpha = 1,2,\ldots,N_{\mathrm{bins}}$)

\begin{align}
    n_{\mathrm{g},\alpha} &= n_{\mathrm{g,tot}} \xi^{\frac{-(\lambda+1)\alpha}{N_{\mathrm{bins}}}} \left( \frac{1-\xi^{\frac{\lambda+1}{N_{\mathrm{bins}}}}}{\xi^{-(\lambda+1)} - 1} \right),\\
    a_{\alpha} &= a_{\mathrm{min}} \xi^{\frac{-\alpha}{N_{\mathrm{bins}}}} \left[ \left(\frac{\lambda+1}{\lambda+3}\right) \left( \frac{1-\xi^{\frac{\lambda+3}{N_{\mathrm{bins}}}}}{1-\xi^{\frac{\lambda+1}{N_{\mathrm{bins}}}}} \right) \right]^{\demi}.
\end{align}
where $\xi = \frac{a_{\mathrm{min}}}{a_{\mathrm{max}}}$.
The total number density of dust $n_{\mathrm{g,tot}}$ is determined by constraining the total grain mass density in the size distribution to be

\begin{equation}
\rho_{\mathrm{g,tot}} = \int_{a_{\mathrm{min}}}^{a_{\mathrm{max}}} \frac{4}{3} \pi \rho_{s} a^{3} \mathcal{C} a^{\lambda} da ~,
\end{equation}
which yields

\begin{equation}
    n_{\mathrm{g,tot}} =\left( \frac{\rho_{\mathrm{g,tot}}}{\frac{4}{3}\pi \rho_s a_{\trm{min}}^3}  \right) \left( \frac{\lambda+4}{\lambda+1} \right) \left( \frac{1-\xi^{-(\lambda+1)}}{\xi^{\lambda+4}-1}  \right) \xi^{\lambda+4} .
\end{equation}

Finally, we constrained the factor $\mathcal{C}$ by choosing the total grain density $\rho_{\mathrm{g,tot}}$ to obtain the same total surface area as a fiducial uniform distribution with a grain size $a_{0}$ \citep[see][]{KunzMouschovias2009}. Because
\begin{equation}
n_{\mathrm{g,tot}} = \frac{a_{\mathrm{max}}^{\lambda+1} \mathcal{C}}{\lambda+1} \left( 1 - \xi^{\lambda+1}\right) ~,
\end{equation}
this yields
\begin{equation}
\rho_{\mathrm{g,tot}} = \rho_{\mathrm{g,tot}}^{\mathrm{fiducial}} \left( \frac{a_{\mathrm{min}}}{a_{0}} \right) \left( \frac{\lambda+3}{\lambda+4} \right) \left( \frac{1-\xi^{-(\lambda+4)}}{1-\xi^{-(\lambda+3)}} \right),
\end{equation}
where, for a dust to gas ratio of 1\%, $\rho_{\mathrm{g,tot}}^{\mathrm{fiducial}} = 0.01~\rho_{\mathrm{n,tot}}$ ($\rho_{\mathrm{n,tot}}$ represents the density of neutral gas).
In the MRN distribution, small grains vastly outnumber the large grains, contributing dominantly to the total surface area.
The total grain density is then $\rho_{\mathrm{g,tot}}^{(\trm{MRN})}~=~0.0341 \rho_{\mathrm{n,tot}}$.

\subsubsection{Grains at high temperature}\label{grains_vap_sect}

\begin{figure}
\begin{center}
\includegraphics[trim = 2cm 2cm 2cm 2cm,width=0.5\textwidth]{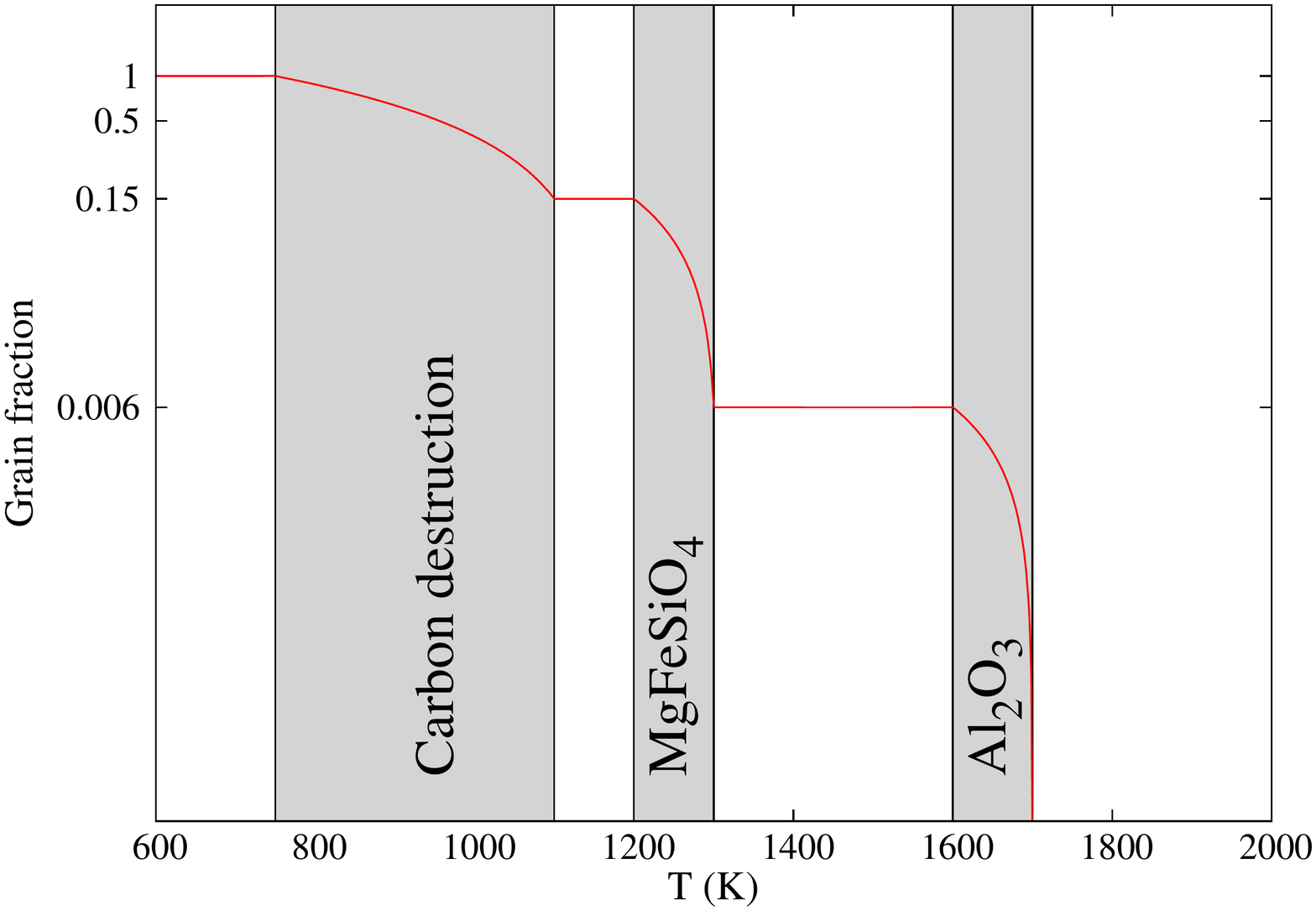}
\caption{Abundance of grains (relative to their value at T $< 750$ K) as a function of temperature. The most abundant species (C, MgFeSiO$_4$ and Al$_2$O$_3$) evaporate during the three represented main destruction stages.}
\label{fracofgrains}
\end{center}
\end{figure}

\paragraph{Grain evaporation}
The three main grain constituents are carbon (essentially amorphous), silicates (here represented by the molecule (MgFe)SiO$_4$) and aluminium oxyde (Al$_2$O$_3$). 
For the sake of simplicity, we assumed that each grain is composed of only one of these three materials, instead of considering a layered structure, as suggested by various studies \protect\citep[e.g.][]{Semenov2003}
The precise evolution of the grain population is a complex issue. Grains evaporate during the first core contraction, before the second collapse. \citet{Lenzuni} proposed two main processes of grain destruction:
 thermal evaporation (destruction directly through thermal vibration), and chemisputtering (reactions between dust and gas). The authors found that carbon evaporates between 750 K and 1100 K, silicates between 1200 K and 1300 K, and aluminium oxides between 1600 K and 1700 K. 
We here assumed that for each material, the quantity of evaporated grains grows linearly with temperature inside the above ranges until total depletion of this type of grain.
Based on the relative fractional abundances of each material calculated from Table 2 in \citet{Lenzuni} C 85\%, (MgFe)SiO$_4$  14.4\%, and Al$_2$O$_3$  0.6\%, 
we obtain the three-step evolution curve displayed in \refig{fracofgrains}.

\paragraph{Thermionic emission}
Thermal agitation on grains induces spontaneous emission of electrons adsorbed on the grain surface \citep{deschturner}. Richardson's law gives the rate of emission
\begin{equation}
  \phi = 4\pi a^2 \lambda_\mathrm{R} \frac{4\pi m_\mathrm{e} (k_\mathrm{B}T)^2}{h^2} \exp \left(\frac{W + \frac{Ze^2}{a}}{k_\mathrm{B}T}\right),
\end{equation}
with $\lambda_\mathrm{R} = 0.5$, $W = 5 \mathrm{eV}$, $Z$ the grain charge and $a$ its radius.

\subsubsection{Grain charges}
Grains can hold several electric charges \citep{DraineSutin}. However, multiply charged grains are weakly abundant in comparison to the singly charged or neutral grains \citep{Sano2004}. Including these grains means handling abundances that range over 20 orders of magnitude, which is numerically difficult to achieve accurately. Furthermore, current analytical models do not correctly describe the charge distribution when grain-grain reactions dominate (see Appendix \ref{Charge_appendix} for more details). In the present paper, we considered grains holding only one electric charge but we rook the grain-grain reactions into account. We acknowledge that multiply charged grains may change our results, and we will address this issue in future work.

\section{Chemistry}\label{chem}

  \subsection{Chemical network}\label{network_sect}

We considered the following elements and their ionised counterparts: H, He, C, O and heavy metallic elements such as Na, Mg, Al, Ca, Fe, Ni, and Si. In conditions typical of molecular clouds and cold neutral medium, H, C and O are primarily found in their molecular forms (H$_2$, CO, O$_2$, H$_2$O, OH). We
 assumed this to be still the case after the second collapse, but we kept in mind that we lack a precise description of the evolution of these molecules at $T>2000~$K.
  The charged particles taken into account are electrons e$^-$, 
H$^+$, He$^+$, C$^+$, H$_\mathrm{3}^+$, molecular ions m$^+$, and metallic ions M$^+$. 
We considered grains of various sizes (see Sect. \ref{grains_size}), either neutral or with an electric charge $\pm e$.
Potassium, sodium, and hydrogen are major  contributors to ionised species in number density at high temperature, because of their low ionisation energy. 
These ionisation reaction{s} \citep[described in][see Appendix \protect\ref{Networkappendix}]{PneumanMitchell} become relevant at $T > 1500~$K 
and are included in our network.
Since Na and K are alkali metal, we assumed that their reactions and associated rate coefficients are the same as for the other metallic ions M$^+$.

Let $\alpha_{ij}$ represent the reactions of ionisation of species $j$ into $i$
\begin{align}
   j \rightarrow i + e^-,
\end{align}
with the ionisation rate of hydrogen molecules $\zeta$. In our context, UV and radionucleides contributions to ionisation rate are negligible compared to cosmic rays that can deeply penetrate dense cores, and we therefore write $\zeta = \zeta_{\mathrm{CR}}$.
 Let $\beta_{ijk}$ represent the reactions between $j$ and $k$ to form $i$
\begin{align}
   j +k \rightarrow i + ,
\end{align}
and $\beta_{*ij}$ (where $*$ denotes any other species that might be present) the reactions between $i$ and $j$ to form another species. 
We note that $\beta_{ijk}=\beta_{ikj}$. We also defined $\gamma_{ij}$ to represent the reactions between $i$ and $j$ to form another species, such as $\beta_{*ij}$, 
 but $\gamma_{ij}$ specifically characterises the destruction of $i$ and $j$ rather than the creation of a given species
\begin{align}
   i + j \rightarrow *. 
\end{align}
 Here $\gamma_{ij}=\gamma_{ji}$.
\noindent We then solved the complete set of equations for each charged species (written in dimensionless form)
\begin{align}
\left\{ 
    \begin{array}{l}
          ... \\
          \frac{d x_i}{\tilde{dt}} = \sum_{j=1}^N \big[ \alpha_{ij} x_j + \frac{n_\mathrm{H}}{2 \zeta} \sum_{k=1}^N \beta_{ijk} x_j x_k - \frac{n_\mathrm{H}}{\zeta} \gamma_{ij} x_j x_i \big] \\
          ...
    \end{array}
\right. \label{base}
\end{align}
where $N$ is the total number of species (both neutrals and charged particles), $n_\mathrm{H}$ is the density of neutrals (here the density of hydrogen molecules), and $x_i$ denotes the fractional abundances of various particles, $x_i = \frac{n_i}{n_\mathrm{H}}$, and $\tilde{dt}=dt \zeta$. 

We considered that neutral abundances are constant, with values taken from \citet{UmebayashiNakano1990}, and we solved for the eight above-mentioned cations, plus electrons and grains.
The reaction rates were taken from the UMIST database \citep{McElroy} for gas species and \citet{KunzMouschovias2009} for the interactions with and between grains. 
More details on the chemical network (the considered reactions, the initial abundances, etc.) are given in Appendix \ref{Networkappendix}.

  \subsection{Numerical method}

Our resolution method is a semi-implicit scheme. It is unconditionally stable and permits either accurate following of the temporal evolution of the solution (using a stringent constraint on the time step) or acceleration of the equilibrium abundances calculations (with a larger time step). In the latter case, we lose precision on the temporal evolution of the network, but the convergence toward equilibrium is unconditional, which is our main interest.

We write $ F(\mathbf{x})_i = \sum_{j=1}^N \big[ \alpha_{ij} x_j + \frac{n_\mathrm{H}}{2 \zeta} \sum_{k=1}^N \beta_{ijk} x_j x_k - \frac{n_\mathrm{H}}{\zeta} \gamma_{ij} x_j x_i \big]$, and Taylor-expand the right-hand side of Eq. (\ref{base}). This yields 

\begin{align}
x_i^{n+1} &= x_i^n + \tilde{dt} F_i ( \mathbf{x}^{n+1}) \nonumber \\
  &= x_i^n + \tilde{dt} \left[ F_i (\mathbf{x}^{n}) + \sum_{j=1}^N \mathbb{J}_{ij}^n \delta x_j + \mathcal{O}(||\mathbf{\delta x}||) \right], 
\end{align}
where $\mathbb{J}$ is the Jacobian matrix, $\mathbf{x} = (x_1,x_2,\dots,x_N)$, and $\delta x_i$ denotes the variation in abundance of the species $i$ between the time steps $n$ and $n+1$. In matrix form, this reads

\begin{align}\label{simplicit}
 (\mathbb{I}-\tilde{dt} \mathbb{J}^n) \,\mathbf{\delta x} =  \tilde{dt} \,\mathbf{F(\mathbf{x^n})},
\end{align}
where $\mathbb{I}$ is the identity matrix, and $\mathbf{F}(\mathbf{x}) = (F_1(\mathbf{x}), \dots, F_n(\mathbf{x}))$.

We limited $\tilde{dt}$ by constraining the maximum allowed relative variation during one time step
\begin{align}
\frac{\Delta \mathbf{x}}{\mathbf{x}} < \epsilon, \label{timestep}
\end{align}
with the control parameter $\epsilon=10^{-2}$.
This permits equilibrium to be rapidly reached and with good precision.
The matrix on the left-hand side of Eq. (\ref{simplicit}) is singular, or numerically very close to singular\footnote{Because of the neutral grains, although it is very close to singular even without them, because of very rare species.}. We used the Singular Value Decomposition (SVD), described in \citet{NumRec} to solve system (\ref{simplicit}).

We imposed $F_i({\mathbf{x})} = 0$ for species whose abundances are more than eight orders of magnitude smaller than the most abundant ones, because their variations have a negligible influence on the chemical evolution, and this avoids unnecessary small time steps. For $T > 1700$ K, we also set up the fractional abundance of grains to $10^{-30}$ instead of to $0$ to avoid numerical problems.

\section{Results}

  \subsection{Fiducial case}

\begin{figure*}
\begin{center}
\includegraphics[width=0.8\textwidth]{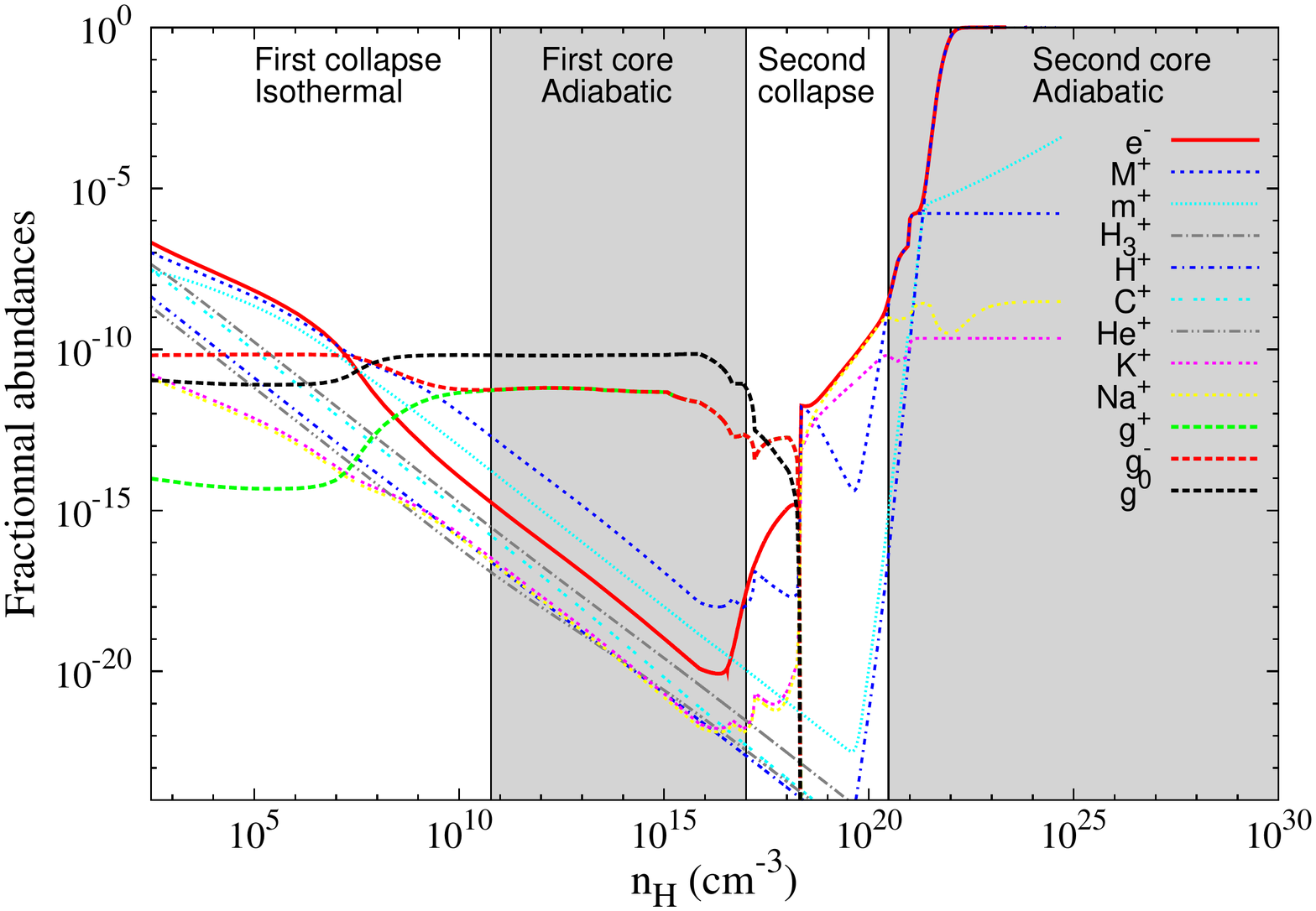}
\caption{Evolution of the fractional abundances for the charged species with density for our fiducial case: 10 bins of grains, barotropic equation of state and a cosmic-ray ionisation rate $\zeta_\mathrm{CR}=10^{-17}$ s$^{-1}$.}
\label{fiducial}
\end{center}
\end{figure*}

  Our fiducial case includes ten bins of grain sizes and a standard cosmic-ray ionisation rate of $10^{-17}$ s$^{-1}$ \citep[as in][]{UmebayashiNakano1990}.
The evolution of the various abundances of charged species as function of density during the global collapse of a prestellar core is presented 
in \refig{fiducial}.

\begin{figure}
\begin{center}
\includegraphics[trim= 2cm 2cm 2cm 2cm, width=0.50\textwidth]{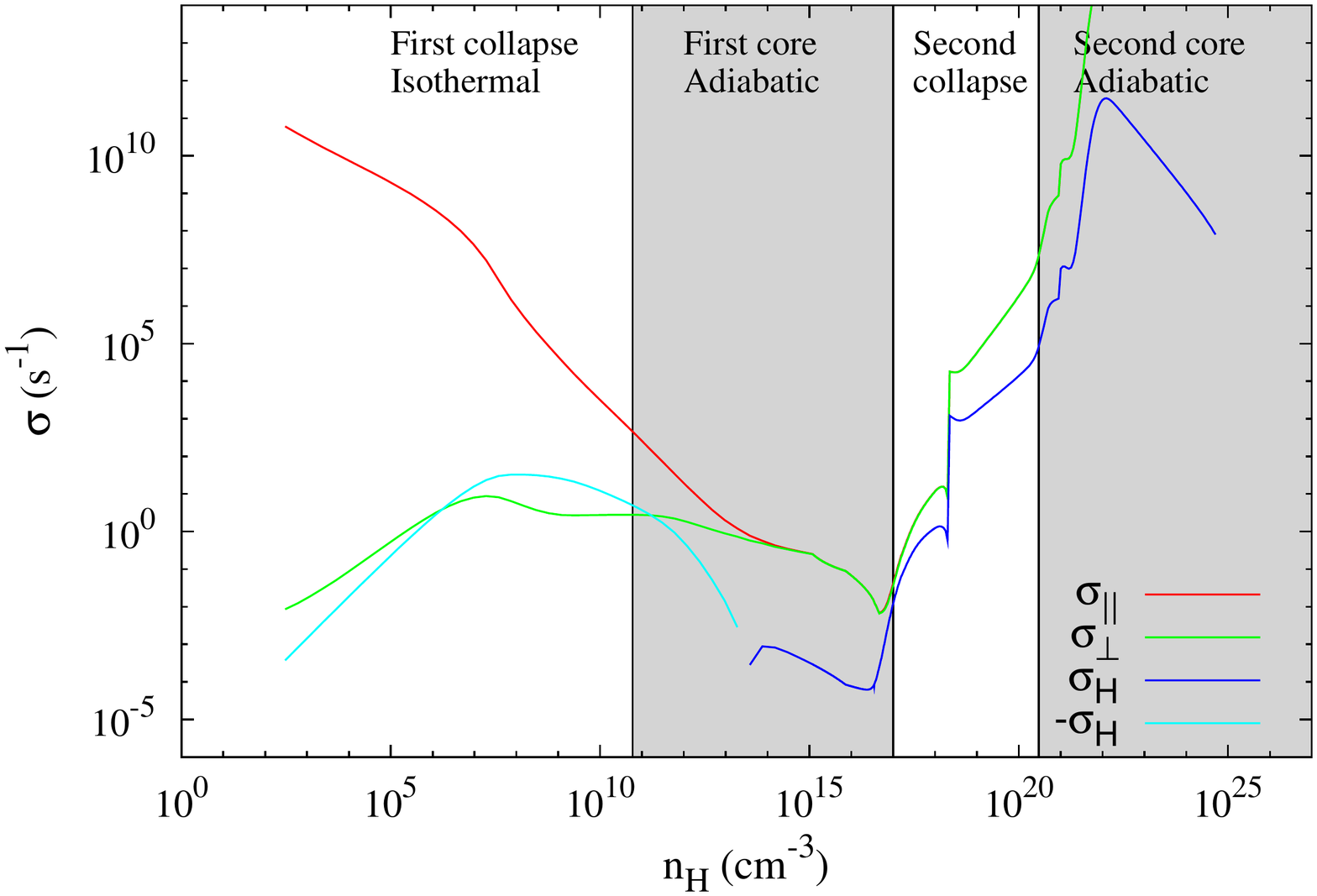}
\caption{Conductivities as a function of density for the fiducial case. Light blue: negative Hall conductivity, dark blue : positive Hall conductivity, red: parallel conductivity, green: perpendicular conductivity.}
\label{conduct}
\end{center}
\end{figure}

\begin{figure}
\begin{center}
\includegraphics[trim= 2cm 2cm 2cm 2cm, width=0.50\textwidth]{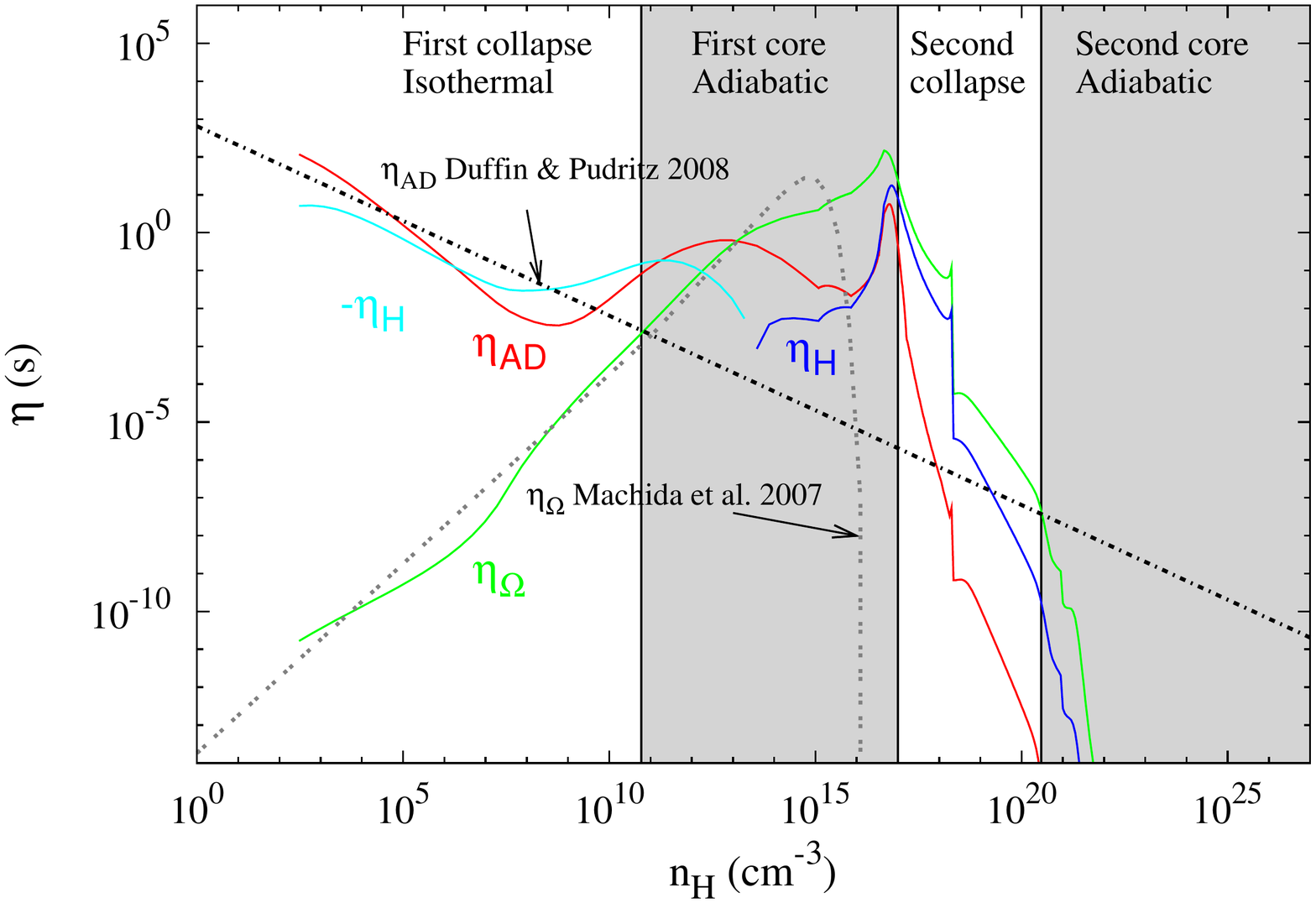}
\caption{Resistivities as a function of density for the fiducial case. Light blue: negative Hall resistivity: dark blue: positive Hall resistivity, red: ambipolar diffusion resistivity, green: Ohm resistivity.}
\label{resist}
\end{center}
\end{figure}

Electrons and M$^+$ are the dominant charged species at low densities. For $n_\mathrm{H} > 10^8$ cm$^{-3}$, grains take over while the abundances 
of all other species decrease by several orders of magnitude. Negatively charged grains are at first more abundant until the neutral grain prevalence for $n_\mathrm{H} > 10^{9}$cm$^{-3}$, eventually becoming the main charge carriers alongside positively charged grains.
At $10^{16}$ cm$^{-3}$, the thermionic emission of grains becomes relevant and releases many electrons. 
Meanwhile, grain evaporation proceeds through the three stages of destruction that are clearly visible until $n_\mathrm{H} = 10^{18}$ cm$^{-3}$.
 This affects the abundances of other species, especially M$^+$, K$^+$, Na$^+$ and e$^-$. Immediately after the complete destruction of the grains, 
the thermal ionisation of K, Na and H become important, and their ionised counterparts become the dominant species along with M$^+$ and e$^-$. 
Eventually, all neutral K, Na and M atoms disappear, leaving H$^+$ as the most abundant ionised species. 
Grain evaporation takes place at the end of the first core phase, and the thermal ionisations essentially take place during the second collapse.
The corresponding conductivities and resistivities are plotted in \refigs{conduct}{resist}. Ohmic and ambipolar resistivities are positive, 
but Hall conductivities and resistivities have negative values at low densities (light blue curves), before becoming positive (dark blue curves). 
For $n_\mathrm{H}~<~10^{15}$ cm$^{-3}$, these figures are qualitatively comparable to Fig. 7 of \citet{Kunz2}. 
We recovered the result of \citet{Wardle1999} concerning the Hall conductivity, becoming comparable to the Pedersen conductivity for 
$10^6~<~n_\mathrm{H}~<~10^{11}$ cm$^{-3}$ for an MRN grain-size distribution, while it is slightly higher in our case. 
There is a peak in resistivities at $n_\mathrm{H}~\approx~10^{18}$ cm$^{-3}$ that is due to grain evaporation, 
where all three resistivities have similar values (but are still dominated by the Ohmic diffusion). 
The peak does not extend over a wide range of densities because the number density of charged species increases again as soon as 
thermal ionisations begin, which drastically decreases the resistivities.
After the peak, resistivity is dominated by the Ohmic and Hall contributions, which remain comparable, until Ohmic resistivity eventually prevails. \refig{resist} also clearly highlights the differences, in the evolution of the various resistivities, between the present calculations and the results of previous studies. For the ambipolar resistivity, \protect\citet{DuffinPudritz} used the simple analytical expression $\eta_\mathrm{AD} \propto \frac{B^2}{n_\mathrm{H}\sqrt{n_\mathrm{H}}}$, so that their resistivity scales as $\frac{1}{n_\mathrm{H}}$ (because we consider $B \propto \sqrt{n_\mathrm{H}}$). Our ambipolar resistivity is close to theirs at low densities, but the two values diverge around $n_\mathrm{H}~=~10^{10}~\cc$. This is due to their assumption that ions are perfectly coupled to the magnetic field, and that the ion-neutral collision time is much shorter than the other characteristic physical times of the system. The authors acknowledge that their model fails for dense regions, with $n_\mathrm{H}~>~10^{10} \cc$. The model of Ohmic resistivity from \citet{Machida_etal07} also matches ours for $n_\mathrm{H}~\le~10^{15} $. The chemical network they considered \citep[][]{nakano2002}, however, does not directly include potassium in this form, neither the present updated reaction rates nor grain evaporation. At low density ($n_\mathrm{H}~\le~10^{[15-16]}~\cc$ for the Ohmic resistivity and $n_\mathrm{H}~\le~10^{10}~\cc$ for ambipolar resistivity), these two commonly used models are then in agreement with our results, but the additional physics included in our work significantly improves the situation for the conditions inside the first and second cores.

  \subsection{Magnetic field variations}

\begin{figure}
\begin{center}
\includegraphics[trim= 2cm 2cm 2cm 2cm, width=0.50\textwidth]{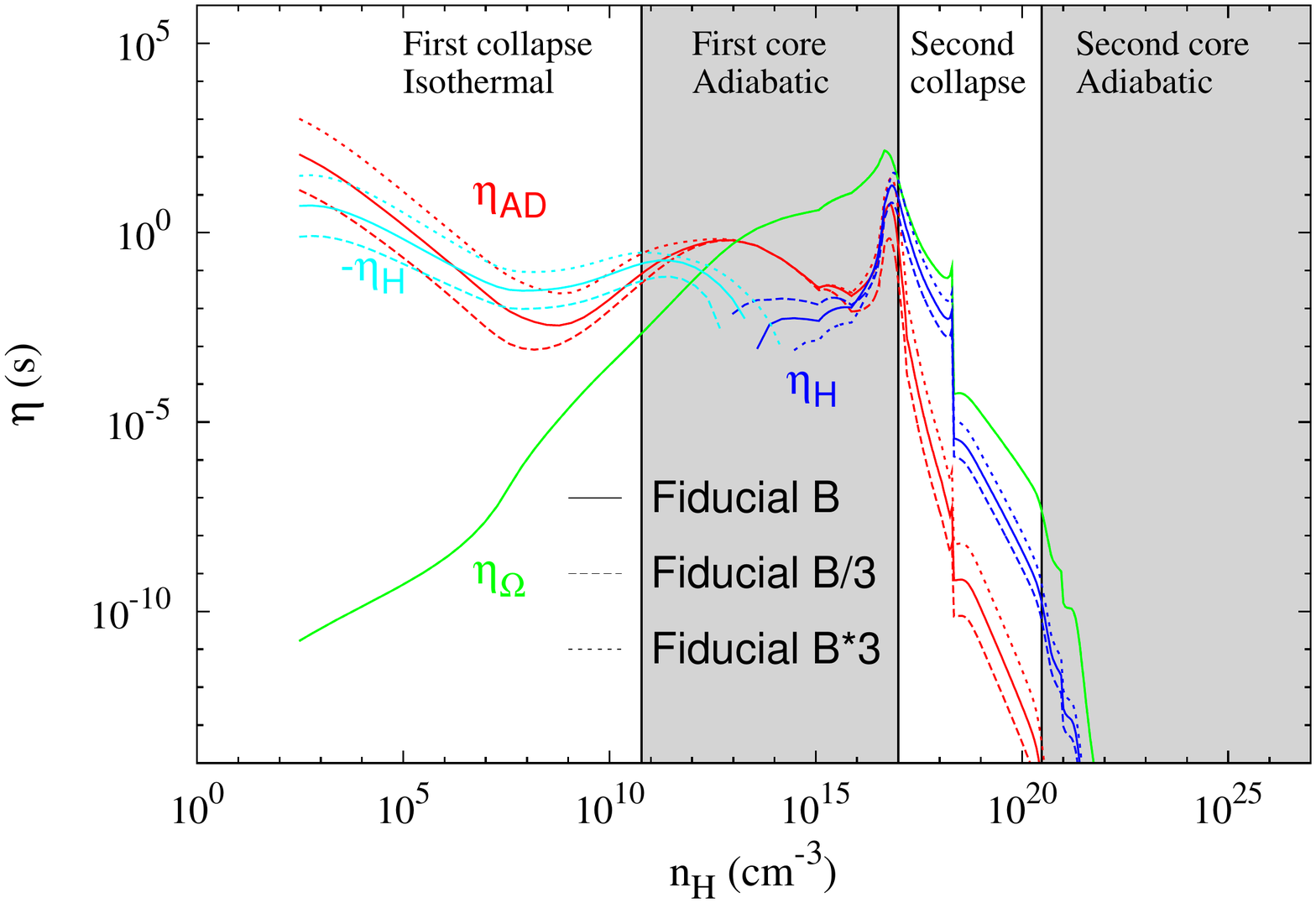}
\caption{Resistivities as a function of density for the fiducial case, and a magnetic field three times higher and lower. Same colour coding as \refig{resist}.}
\label{mag_change}
\end{center}
\end{figure}

The magnetic field only influences the resistivities and the conductivities and has no effect on the chemistry. \refig{mag_change} shows the resistivities for the fiducial case and for magnetic fields three times lower and higher. Hall and ambipolar resistivities are slightly shifted, while the Ohmic resistivity is of course not affected because it does not depend on $B$. Varying the magnetic field has the strongest effect on the first collapse density range. The density at which the Hall resistivity changes sign during the first core contraction is also shifted, and 
the ratio between the three resistivities is strongly affected, especially in the $10^{16}-10^{18}$ cm$^{-3}$ density range. Because the resistivities were computed on the fly with the magnetic field of the simulation, this variation will influence the gravitational collapse. As the Hall effect is strongly sign dependent, the formation of some structures, like the first Larson core or the protoplanetary disk, may be delayed or made impossible \citep{Tsukamoto2015,Wurster2015}.

  \subsection{Effect of grain evaporation}

\begin{figure}
\begin{center}
\includegraphics[trim= 2cm 2cm 2cm 2cm, width=0.50\textwidth]{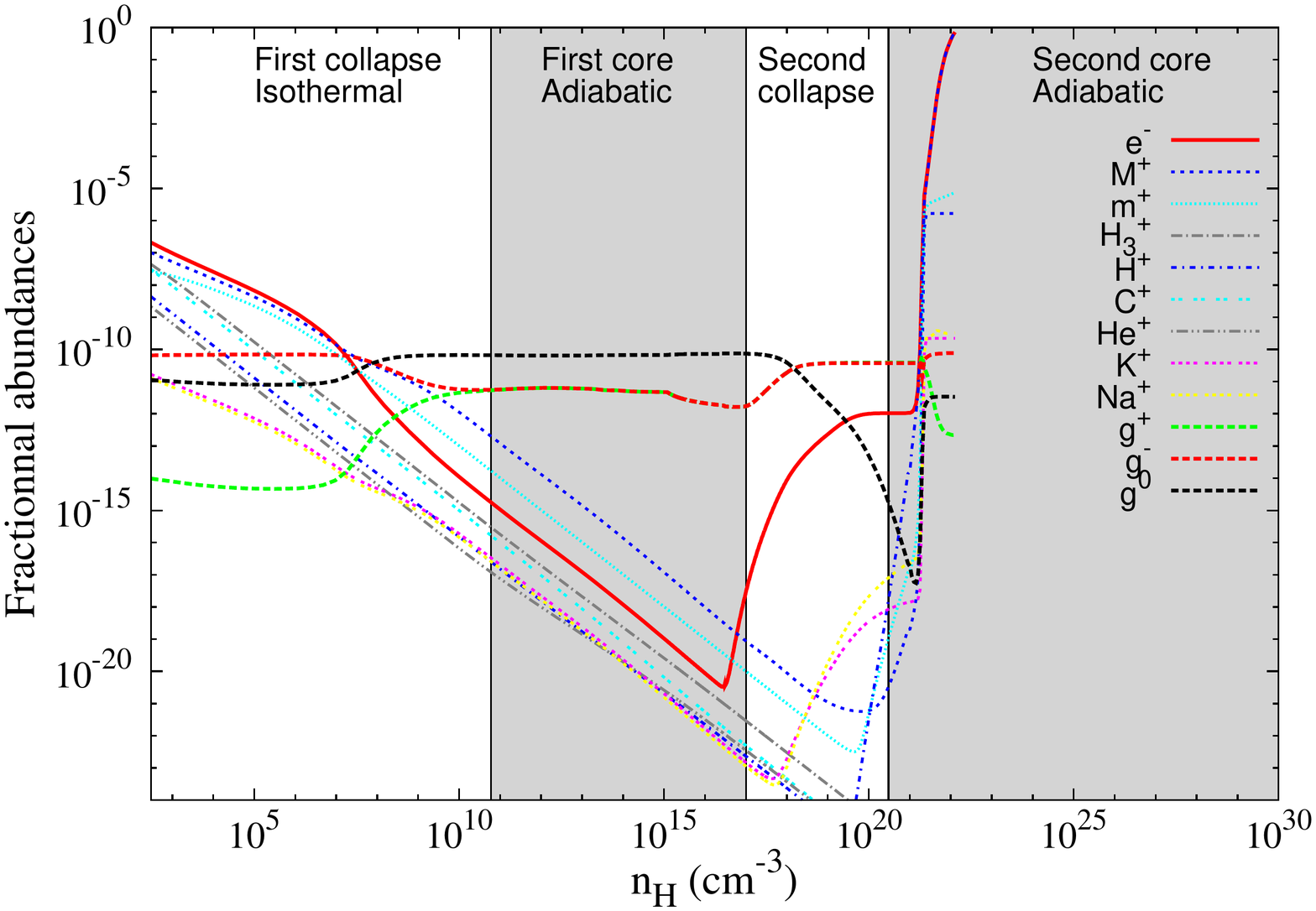}
\caption{Fractional abundances, using ten bins, without grain evaporation.}
\label{novap_nh}
\end{center}
\end{figure}

\begin{figure}
\begin{center}
\includegraphics[trim= 2cm 2cm 2cm 2cm, width=0.50\textwidth]{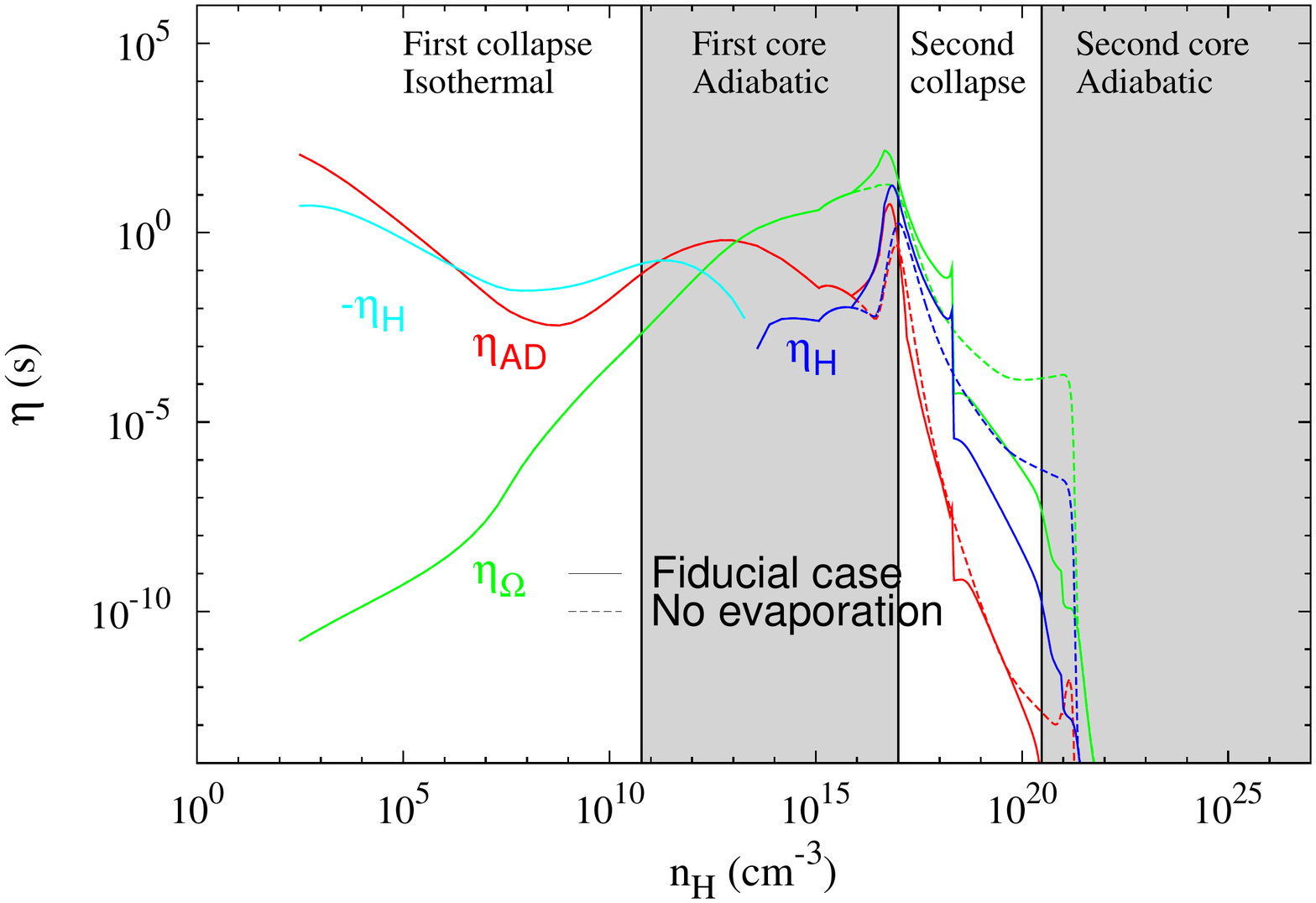}
\caption{Comparison of the resistivities with and without grain evaporation. Same colour coding as in \refig{resist}. Dashed lines represent the absence of evaporation, whereas solid lines stand for the fiducial case.}
\label{novap_eta}
\end{center}
\end{figure}

  \refig{novap_nh} presents the same calculations as for the fiducial case, but without grain evaporation. There are substantial changes for $n_\mathrm{H} > 10^{16}$ cm$^{-3}$, at the onset of grain destruction.
  The abundances of both electrons and metallic ions increase as a result of the thermionic emission of the grains and the thermal ionisations. 
 They are the dominant species along with negatively charged grains, which are more prone to form because of the large number of free electrons released by K, 
Na and H atoms. Consequently, the abundances of neutral and positive grains both drop, while the abundances of other species remain very small. 
\refig{novap_eta} shows the effect that removing grain evaporation has on the resistivities. At high densities ($>10^{20}$~cm$^{-3}$) 
the resistivities still decrease because the number of ionised particles increases in the medium. 
In the density range typical of the second collapse and the second Larson core, 
$10^{15} < n_\mathrm{H} < 10^{22}\trm{cm}^{-3}$, the Hall and the Ohmic resistivites show significant differences compared to the fiducial case. 
The Hall resistivity is first lower and then higher than the reference case, while the Ohmic resistivity is tremendously higher in this density range.
At $n_\mathrm{H} \approx 10^{22}\trm{cm}^{-3}$, the excessive abundance of electrons in the medium cancels out the difference between the two cases. 
An accurate description of grain evaporation is therefore mandatory to study the possible transformation of the first core into a disk around the second Larson core \citep[as  described in][]{Machida_etal06}.

  \subsection{Cosmic-ray ionisation rates} \label{ionisation_sect}

Cosmic-rays (hereafter CR) propagation along field lines is affected by two competing effects: magnetic focusing, which increases the ionisation rate, and magnetic mirroring, which prevents CRs from reaching deep parts of the cloud. \citet{PadovaniGalli2011,PadovaniHennebelleGalli2013,2014A&A...571A..33P} found that mirroring always dominates focusing for a field topology obtained by ideal-MHD simulations of collapsing rotating clouds and that the ionisation rate could vary by up to a factor $50$, depending on the mass-to-flux ratio and the magnetic flux tube considered. In addition, CR are partly absorbed by the dense medium, which lowers the CR ionisation rate \citep{Padovani2009}.

\begin{figure}
    \begin{center}
        \includegraphics[trim= 2cm 2cm 2cm 2cm, width=0.50\textwidth]{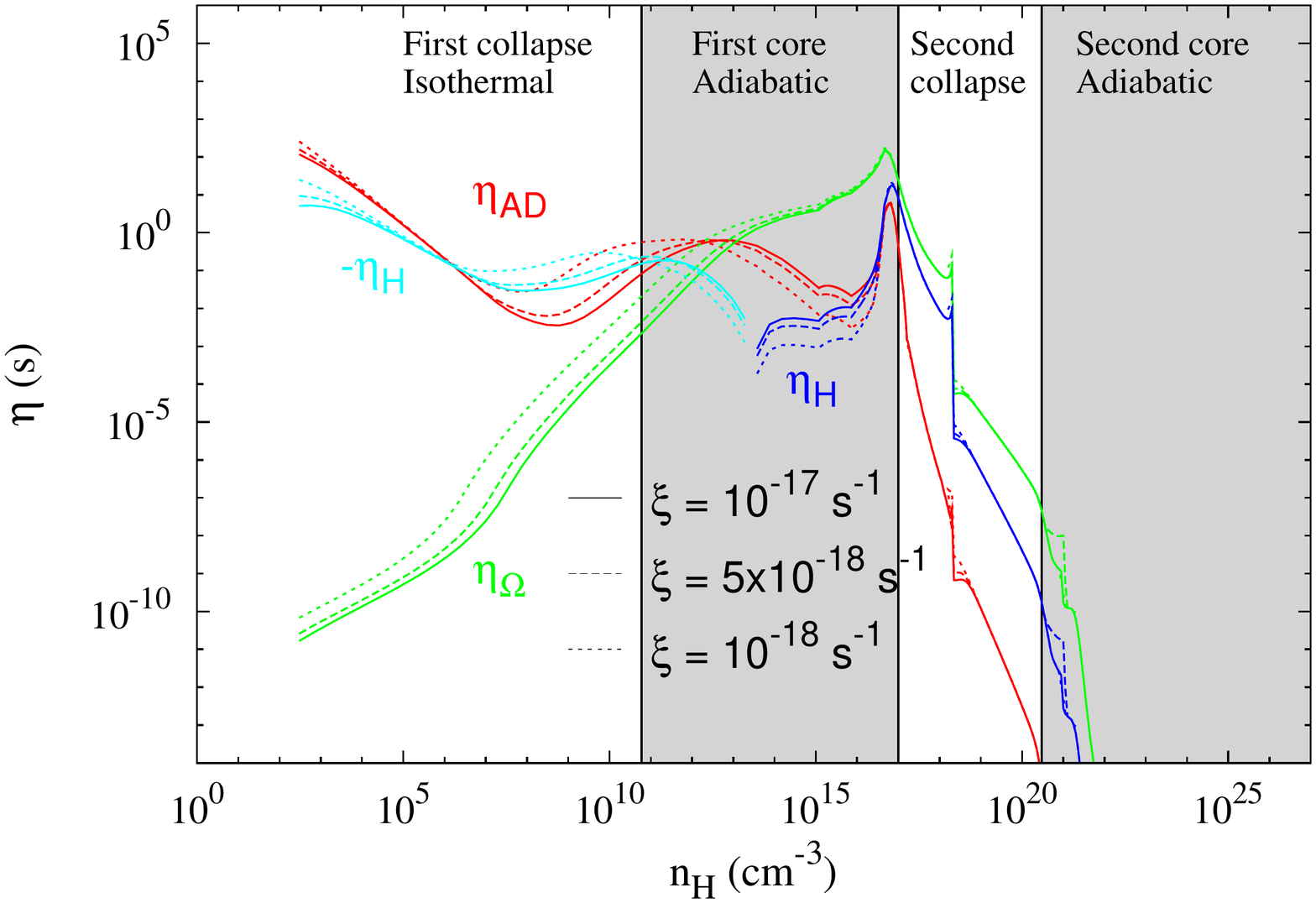}
        \caption{Resistivities for different cosmic-ray ionisation rates, with the barotropic EOS. Solid lines: fiducial value $\zeta_{\mathrm{CR}_{\trm{ref}}} = 1\times 10^{-17} \mathrm{s}^{-1}$. Dashed lines: $\zeta_\mathrm{CR} = \frac{\zeta_{\mathrm{CR}_\trm{ref}}}{2} = 5\times 10^{-18} \mathrm{s}^{-1}$. Dotted lines: $\zeta_\mathrm{CR} = \frac{\zeta_{\mathrm{CR}_\trm{ref}}}{10} = 1\times 10^{-18} \mathrm{s}^{-1}$. Same colour coding as \refig{resist}}
        \label{ionis_try}
    \end{center}
\end{figure}

To quantify the effect of these uncertainties on the CR ionisation rate in a protostellar environment with a complex magnetic field topology (twisted field lines, misalignement, turbulence, etc.), we computed abundances and resistivities for two CR values in addition to our standard case, namely $\zeta_\mathrm{CR} = 5\times 10^{-18} \, \mathrm{s}^{-1}$ and $\zeta_\mathrm{CR} = 1\times 10^{-18} \, \mathrm{s}^{-1}$. The results are shown \refig{ionis_try} (we do not show the abundances because they are similar to the fiducial case).
 The Hall and ambipolar diffusion contributions are alternatingly increase and decrease. 
In comparison, our fiducial value $\zeta_\mathrm{CR} = 1\times 10^{-17} \, \mathrm{s}^{-1}$ yields about an Ohmic dissipation that is about an order of magnitude smaller 
 before the resistivities decrease at $n_\mathrm{H} \approx 10^{17}$~cm$^{-3}$. 
Minor deviations are visible in the second collapse density range, but the overwhelming contribution of electrons and hydrogen compensates for most of the variations.

  \subsection{Grain-size distribution}\label{sizedistrib}

Most of the grain surface is due to the smallest grains, therefore a large enough number of grain bins is necessary to properly evaluate of the final conductivity.
Similar as \citet{KunzMouschovias2009}, we found that five size bins seem to be sufficient to reach a relative precision of the order of $1\%$ for every considered species. In this part, our reference case includes ten size bins. The error on the abundances was calculated as $\frac{|| \mathbf{x}_{N_\mathrm{bins}} - \mathbf{x}_\mathrm{ref}||}{||\mathbf{x}_\mathrm{ref}||}$, where $||\ ||$ is the $\mathrm{L}_2$ norm of the abundance vector $\mathbf{x}$, $\mathbf{x}_{N_\mathrm{bins}}$ is the abundance vector for the number of bins $N_\mathrm{bins}$ considered, and $\mathbf{x}_\mathrm{ref}$ is the abundance vector for the ten bin case. The relative error on the least abundant gas molecules becomes fairly high when grains start to evaporate at $10^{17}~$cm$^{-3}$, but this is inconsequential since these species hardly contribute to the resistivity at this stage. Five bin abundances clearly yield a smaller error than the ten bins case, which shows that our fiducial calculations have a good precision.
The resistivities for different numbers of bins (one, five, and ten) are plotted in \refig{resist_bins}. 
When only one bin is considered, the resistivities are shifted up and down by about one order of magnitude and are strongly overestimated in 
the $10^{12} - 10^{17}~$cm$^{-3}$ density range. However, the five and ten bins cases are extremely close to each other.
 A large enough number of bins is thus necessary to properly describe this highly dynamic phase of the prestellar core evolution. 

\begin{figure}
\begin{center}
\includegraphics[trim= 2cm 2cm 2cm 2cm, width=0.50\textwidth]{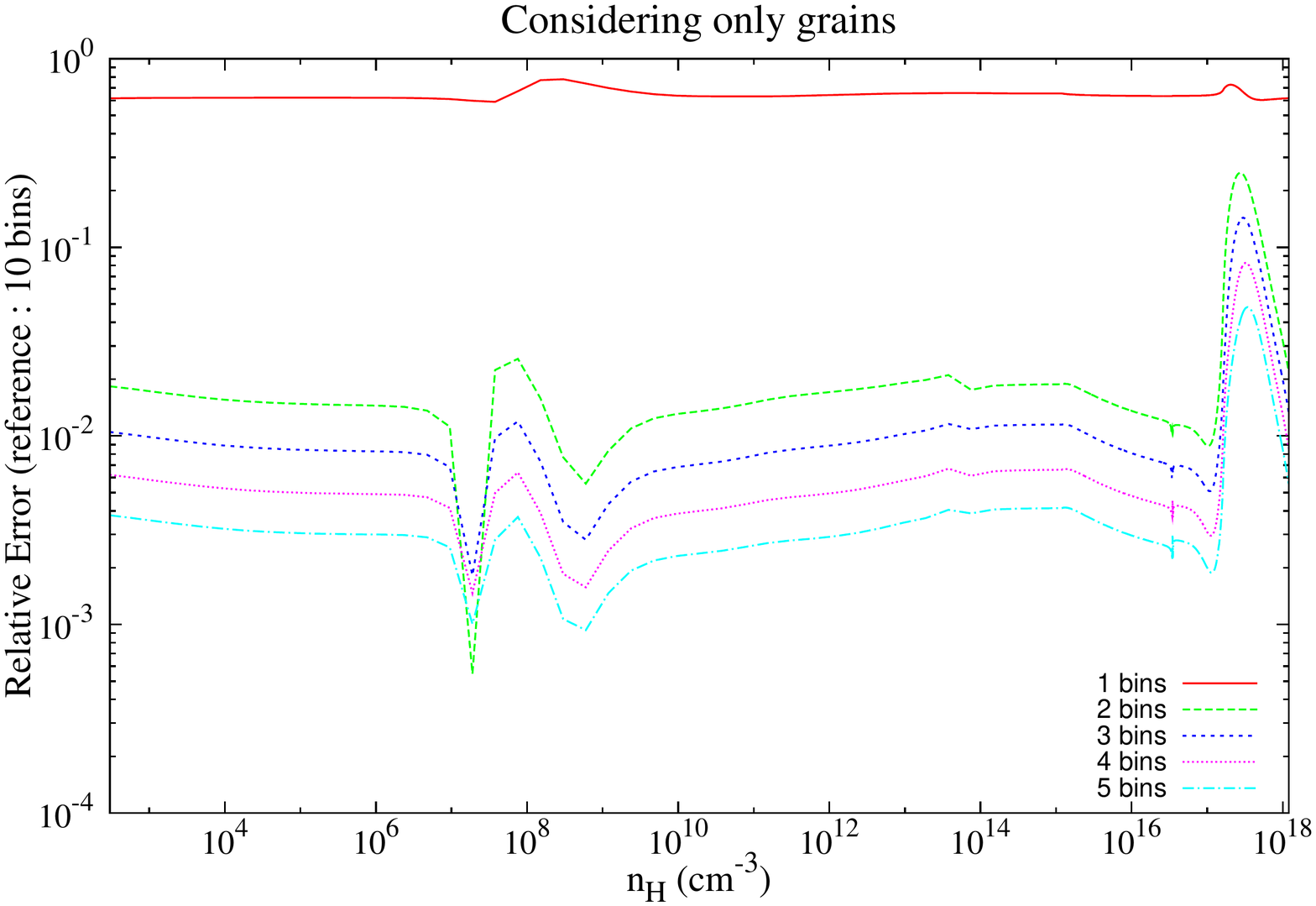}
\includegraphics[trim= 2cm 2cm 2cm 2cm, width=0.50\textwidth]{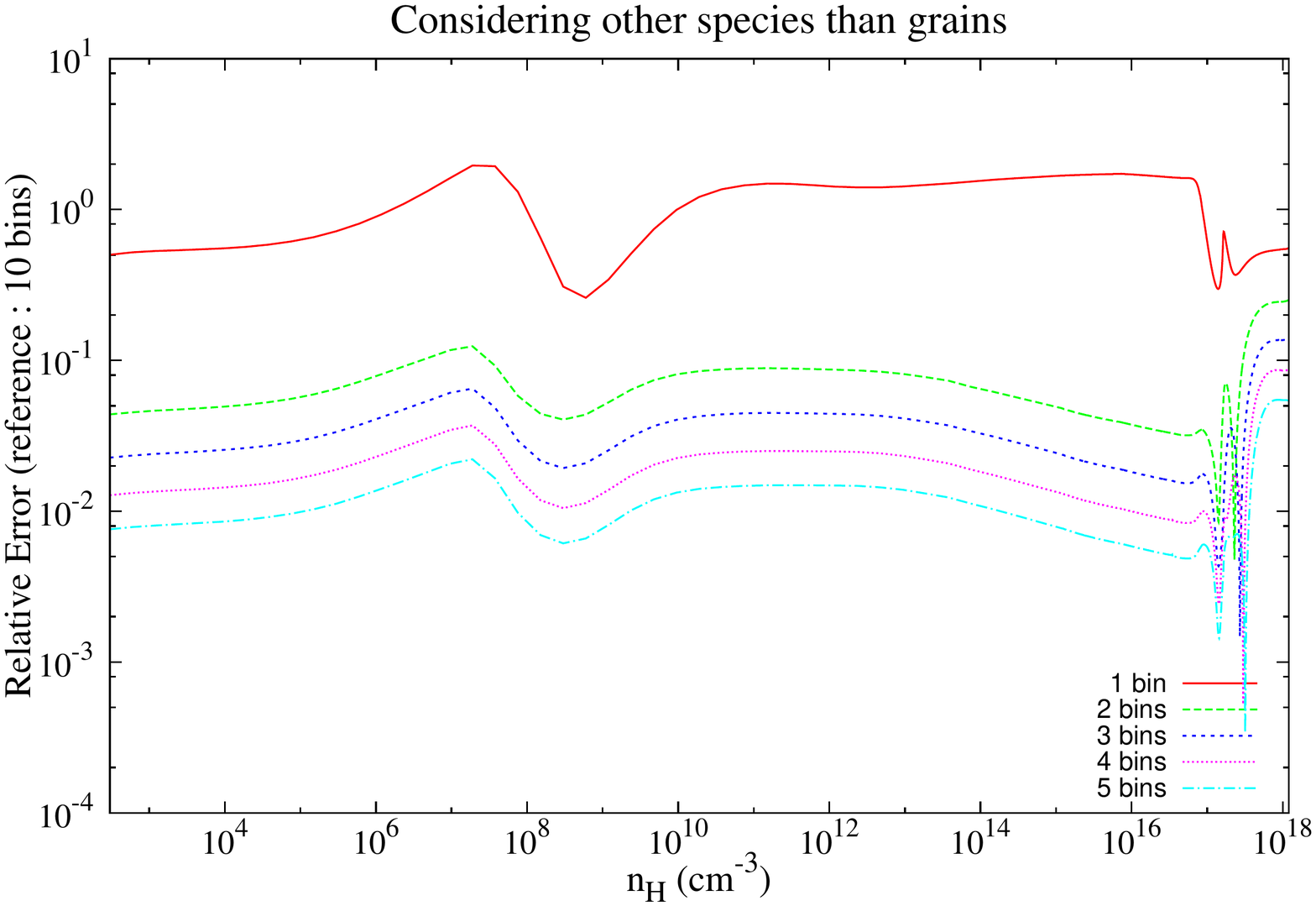}
\caption{Relative error in the species abundances ($=\frac{||\mathbf{x_{N_\mathrm{bins}}}-\mathbf{x_{ref}}||}{||\mathbf{x_{ref}}||}$) for different numbers of size bins (relatively to the ten bins results) for grains only (top panel), and for other species (bottom panel) in the density range before grain evaporation.}
\label{precisiongrains}
\end{center}
\end{figure}

\begin{figure}
\begin{center}
\includegraphics[trim= 2cm 2cm 2cm 2cm, width=0.50\textwidth]{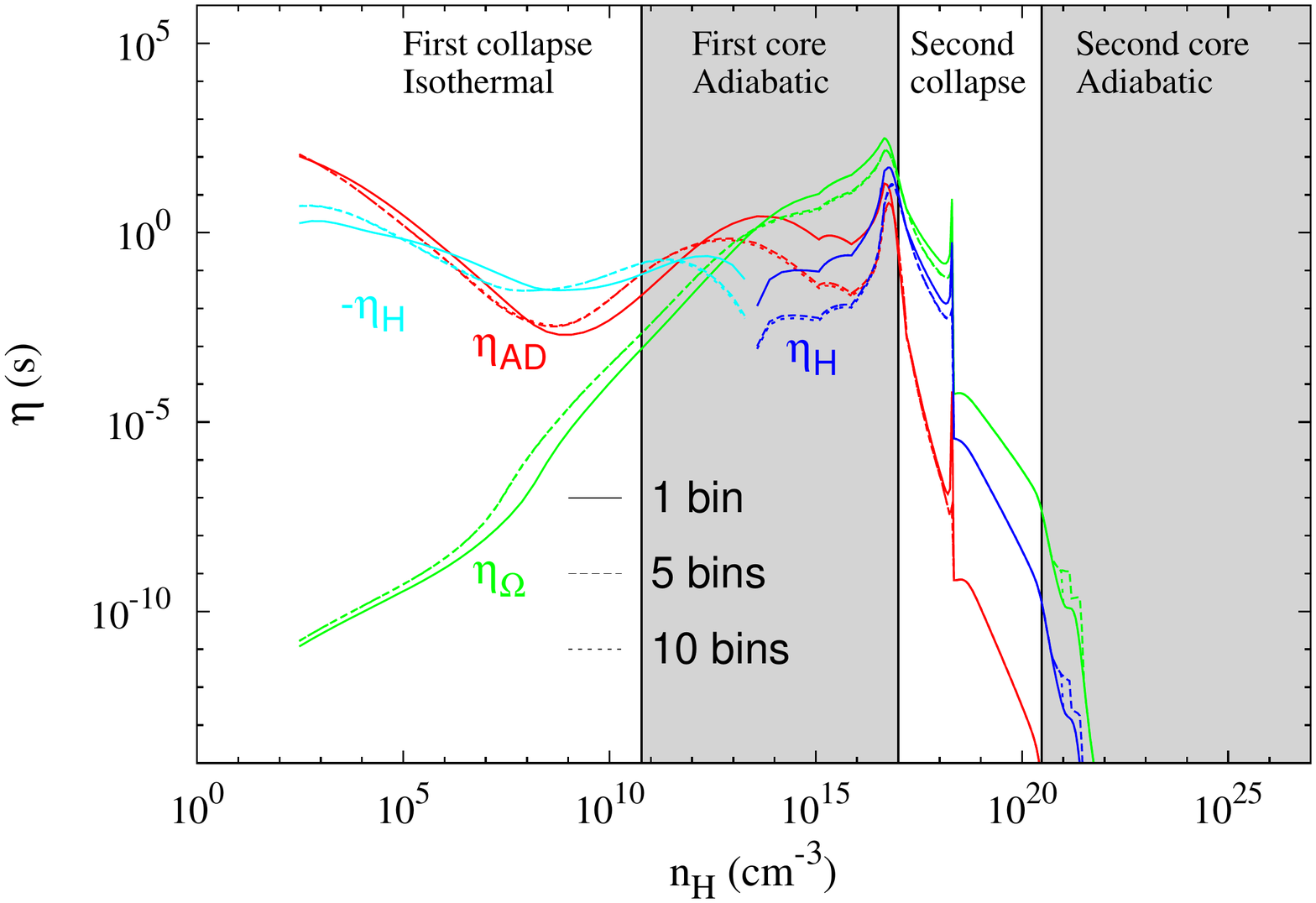}
\caption{Resistivities for different numbers of size bins. Solid line: ten bin fiducial case, dotted line: five bins, dash-dotted line: one bin. Same colour coding as \refig{resist}.}
\label{resist_bins}
\end{center}
\end{figure}

\begin{figure}
\begin{center}
\includegraphics[trim= 2cm 2cm 2cm 2cm, width=0.5\textwidth]{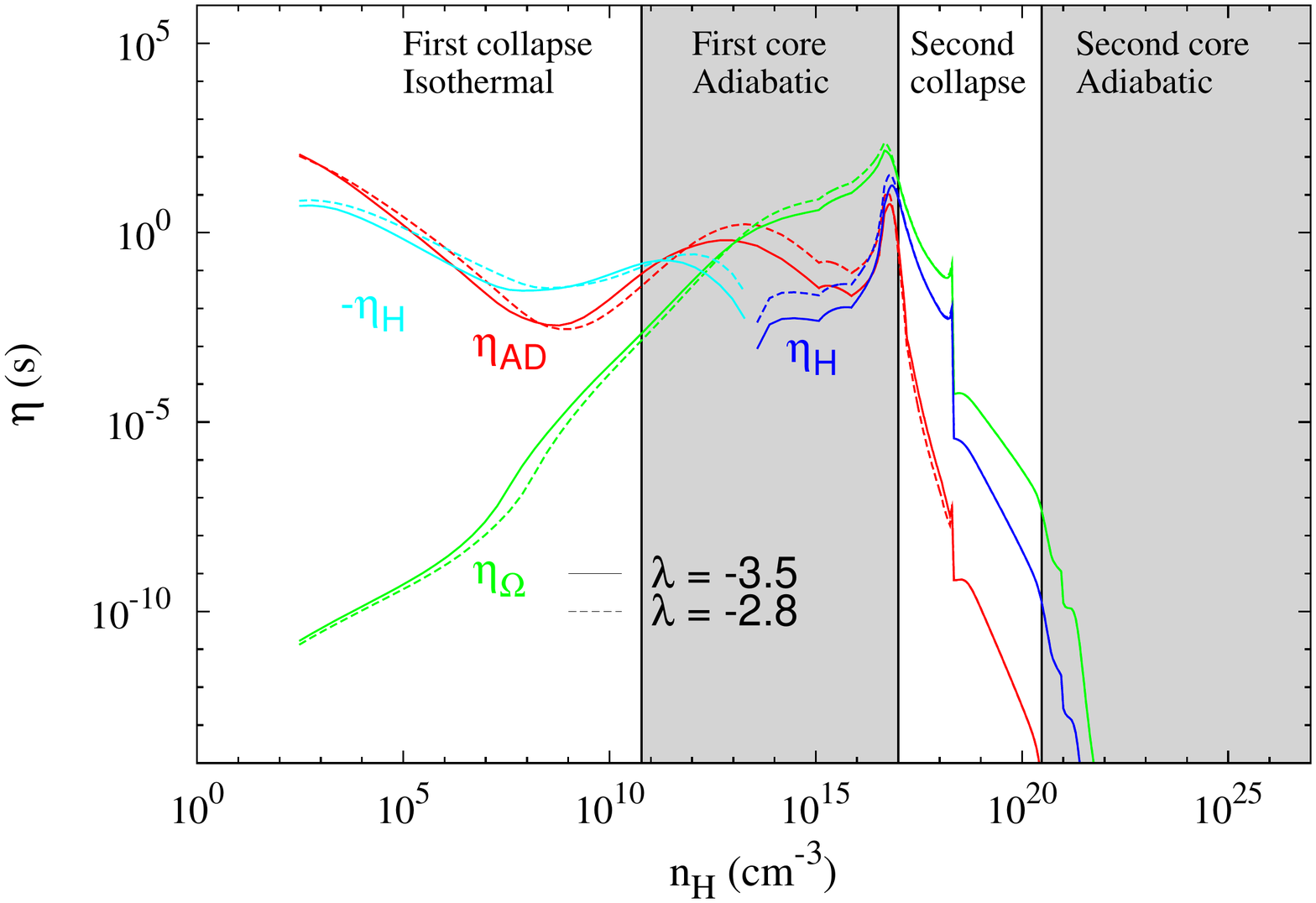}
\caption{Ambipolar diffusion resistivity for the MRN $\lambda=-3.5$ (full line) and the \citet{Compiegne} $\lambda=-2.8$ (dashed line) distributions, with the same (our) barotropic EOS.}
\label{bins_inv_fig}
\end{center}
\end{figure}

Even though the MRN grain-size distribution is reasonable when considering interstellar dust, it becomes of questionable validity for denser media such as molecular clouds or Larson cores, for which a precise grain-size distribution is still lacking.
We examined the effect of this uncertainty by conducting calculations with another distribution: $\lambda = -2.8$ 
(instead of $\lambda = -3.5$ for the MRN distribution), as suggested by \citet{Compiegne} for amorphous carbon grains between radii of$\sim 5$ 
and few hundred nanometers. The result for the resistivities is shown in \refig{bins_inv_fig}. 
Despite the similarity of the comportment of the resistivities between $10^5$ cm$^{-3}$ and $10^{17}$ cm$^{-3}$, 
there is a slight difference in the partitioning for the dominant effect. For example, around $n_\mathrm{H} = 10^{14}$ cm$^{-3}$, 
Ohmic diffusion takes the lead upon the ambipolar diffusion for a density ten times higher than the MRN distribution. 
This density range is typical of protostellar disks, which means that the respective evolution of these latter might be quite different because the size distributions may vary from one to another, with effects difficult to control (grain coalescence, premature destruction, etc.). 
Several authors proposed their own size distribution. \citet{Compiegne} also assumed a lognormal distribution for the smallest grains (PAH and small amorphous carbon) and a power law with an exponential cut-off for larger amorphous carbon grains and amorphous silicate grains to reproduce observed emission and extinction spectra (see their Fig. 2).
 
  \subsection{Non-equilibrium chemistry}

\begin{figure}
    \begin{center}
        \includegraphics[trim= 1cm 2cm 1cm 2cm,width=0.49\textwidth]{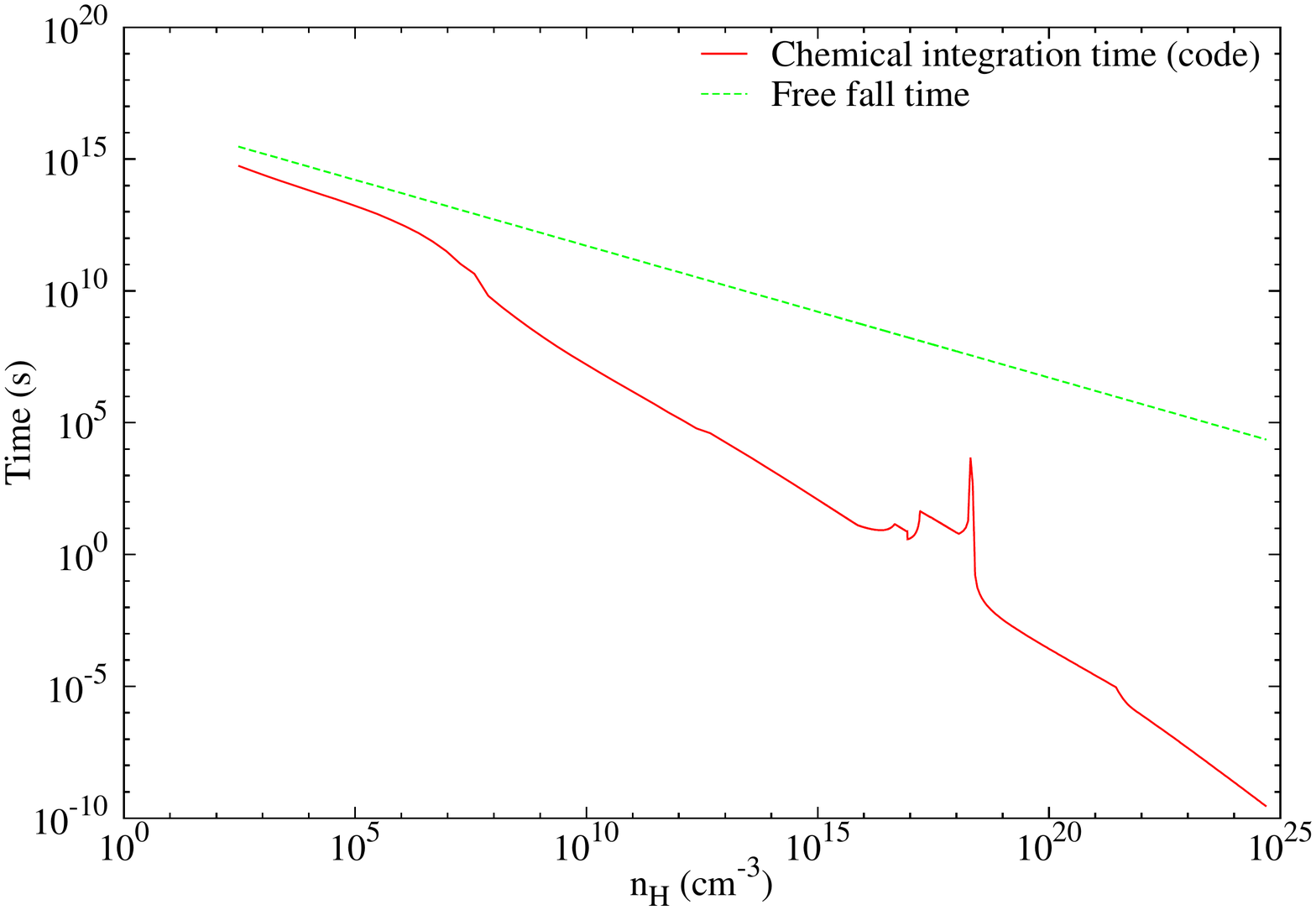}
        \caption{Typical chemical equilibrium timescale (solid red) and dynamical timescale (dash green) as a function of density.}
        \label{times}
    \end{center}
\end{figure}

Until now, all the species abundances were calculated at chemical equilibrium. In real situations, the chemical reaction timescale could be greater than the dynamical (collapse) one. In that case, the environment conditions (density, temperature, etc.) can change significantly before chemical equilibrium is reached. 
We explored this possibility by comparing the timescale required for our chemical network to reach equilibrium with the typical free-fall time for a self-gravitating spherical cloud,
$t_\mathrm{ff} = \sqrt{\frac{3\pi}{32G\bar{\rho}}}$ (with $\bar{\rho}$ the mean density of the cloud). The results are portrayed in \refig{times}. Clearly, the free-fall time is far longer than the chemical equilibrium time at all densities. Furthermore, the above free-fall time estimate is likely to be underestimated since real clouds are additionally supported by thermal, turbulent and magnetic pressures. Therefore, assuming chemical equilibrium seems to be valid in the context of collapsing prestellar cores, with the limitation that we did not consider the history of the chemical species during the collapse. Particles may be transported from one specific environment to another (e.g. from the core region to the outflow then to the disk). The only way to properly account for this complicated pattern is to calculate non-equilibrium chemical reactions on the fly during the dynamical collapse. Additionally the flow of the fluid changes the dynamics and statistics of the collisions, which results in chemical transformations and should therefore also be taken into account. This task remains computationally prohibitive for now.

\section{Conclusion}

We have developed a solver that computes a detailed network of the main chemical reactions relevant to the first and second collapse of prestellar cores. The network is based on the work of \citet{UmebayashiNakano1990} but extends this study significantly by updating the reaction rates and including the effects of dust evaporation, thermal ionisation of potassium, and exploring various cosmic-ray ionisation rates. We also used a distribution of grain sizes, the MRN distribution, and explored the effect of the number of size-bins on the results. The solver yields the chemical equilibrium abundances of the various neutral and ionised species\footnote{It can also be used to derive out-of-equilibrium abundances of given species during the collapse, if necessary.}. The various abundances were used to calculate the non-ideal MHD resistivities, namely Ohmic, Hall and ambipolar resistivities, during the collapse, using a barotropic EOS to reproduce the typical density-temperature conditions. The resistivities determine the dynamics of the first and second collapse, and thus the properties of the first and second Larson cores.

Above a temperature $T \sim 500~$K , the effects of thermionic emission, grain evaporation and thermal ionisation become preponderant. An accurate description of these processes is mandatory to properly characterise the collapse because they occur during the first core contraction and influence the initial conditions of the second collapse. Dust destruction has a double effect on the collapse. First, it modifies the opacity of the medium \protect\citep[see e.g.][]{Lenzuni}, which regulates the radiative cooling of the system. Evaporation increases the efficiency of this process, and the collapse accelerates as the gas is less thermally supported. Second, it affects the various resistivities of the non-ideal MHD terms with direct effect on magnetic field topology and a diminution of the magnetic braking.

 We did not include the molecular dissociation of elements in the chemical network, in particular the dissociation of H$_2$ at 2000 K, which leads to the second collapse. However, we do 
not expect the resistivities to be significantly affected by this process because H$^+$ ions and electrons are the dominant charged species at these temperatures. As mentioned in the text, our grain model is simplified because we assumed that each grains was composed of only one material. A more realistic structure containing several layers of different materials that evaporate one after the other will be included in future work.

In addition to the general outcome of these calculations and their effect on prestellar core evolution, we wish to highlight the following points
\begin{itemize}
\item At least five bins for the grain-size distribution are necessary to reliably determine of the resistivities. As discussed in Sect. \ref{sizedistrib}, a precise knowledge of the grain-size distribution would certainly improve the reliability of the results.
\item As shown in Sect. \ref{ionisation_sect}, resistivities change by up to two orders of magnitude when ionisation rates vary by a factor 10. This highlights the importance of shielding against cosmic rays in collapsing cores \citep{Padovani2014}.
\item In the temperature range 750 K $\lesssim T \lesssim 1700$ K, dust grains evaporate. This evaporation has tremendous consequences on the various chemical abundances and thus on the resistivities, since grains are the main contributors to the resistivities at these temperatures.
\item Around 1500 K and above, at which thermal ionisation of metallic ions and hydrogen occur, grains have been entirely destroyed and H$^+$ and electrons become the main charge carriers, which causes the resistivities to drop even further.
\item Our chemical integration time is always shorter than the free-fall time. We can therefore assume equilibrium chemistry, which is less demanding than non-equilibrium chemistry, especially during hydrodynamics simulations.
\end{itemize}
This solver allows us to compute a large multi-dimensional multi-species equilibrium abundance table for a wide range in temperatures, densities and ionisation rates.
This table can be used during simulations of the first and second collapse of prestellar cores, allowing a consistent dynamical-chemical description of this process. The table can be downloaded at 
https://bitbucket.org/pmarchan/chemistry.

\begin{acknowledgements}
The research leading to these results has received funding from the European Research Council under the European Community's Seventh Framework Programme (FP7/2007-2013 Grant Agreement no. 247060).
BC gratefully acknowledges support from the French ANR Retour Postdoc program (ANR-11-PDOC-0031).
NV is supported by the European Commission through the Horizon 2020 Marie Sk{\l}odowska-Curie Actions Individual Fellowship 2014 programme (Grant Agreement no. 659706).
We acknowledge financial support from the ``Programme National de Physique Stellaire'' (PNPS) of CNRS/INSU, France.
We thank T.~Grassi for useful discussions during the writing of this paper. Finally, we would also like to thank the referee for very insightful comments that have greatly helped to improve the completeness of this work.
\end{acknowledgements}

\begin{appendix}

\section{Details of the chemical network}\label{Networkappendix}

Table \ref{reactions} gives a list of all the reactions implemented in the code, taken from \citet{UmebayashiNakano1990}. The reaction rates are taken from the UMIST Database \citep{McElroy}. The $\bar{\alpha}$, $\bar{\beta}$, and $\bar{\gamma}$ coefficients are defined as
\begin{align*}
k=\bar{\alpha} \left(\frac{T}{300}\right)^{\bar{\beta}} e^{-\frac{\bar{\gamma}}{T}},
a=1
\end{align*}
with $k$ the reaction rate. The units of $k$ are cm$^3$s$^{-1}$ for reactions between species and s$^{-1}$ for ionisation and photo-reactions with cosmic rays.
M stands for metals such as Mg, Al, Ca and Fe. Following \citet{UmebayashiNakano1990}, 
we used for all these elements the typical coefficient rates of Mg, which is the most abundant species. 
m stands for molecules that can be ionised and are represented by HCO$^+$.

\begin{table}
\caption{Chemical reactions and coefficient rates of the chemical network.}
\label{reactions}
\centering
\begin{tabular}{lrrr}
\hline\hline
Reaction & $\bar{\alpha}$ & $\bar{\beta}$ & $\bar{\gamma}$\\
\hline
H$^+$ + O $\rightarrow$ H + O$^+$    &    $6.86 \times 10^{-10}$    &    $0.26$    &   $0$\\
H$^+$ + O$_2$ $\rightarrow$ H + O$_2^+$    &    $2.00 \times 10^{-9}$    &    $0.00$   &   $0$\\
H$^+$ + M $\rightarrow$ H + M$^+$    &    $1.10 \times 10^{-9}$    &    $0.00$   &   $0$\\
He$^+$ + H$_2$ $\rightarrow$ He + H$^+$ + H    &    $3.70 \times 10^{-14}$    &    $0.00$   &    $35$\\
He$^+$ + CO $\rightarrow$ He + C$^+$ + O   &    $1.60 \times 10^{-9}$    &    $0.00$    &   $0$\\
He$^+$ + O$_2$ $\rightarrow$ He + O$^+$ + O   &    $1.10 \times 10^{-9}$    &    $0.00$    &   $0$\\
H$_3^+$ + CO $\rightarrow$ H$_2$ + HCO$^+$   &    $1.36 \times 10^{-9}$    &    $-0.14$    &   $0$\\
H$_3^+$ + O $\rightarrow$ H$_2$ + OH$^+$   &    $7.98 \times 10^{-10}$    &    $-0.16$    &   $0$\\
H$_3^+$ + O$_2$ $\rightarrow$ H$_2$ + O$_2$H$^+$   &    $9.30 \times 10^{-10}$    &    $0.00$    &   $0$\\
H$_3^+$ + M $\rightarrow$ H$_2$ + H + M$^+$   &    $1.10 \times 10^{-9}$    &    $0.00$    &   $0$\\
C$^+$ + H$_2$ $\rightarrow$ CH$_2^+$ + $h\nu$    &    $2.00 \times 10^{-16}$    &    $0.00$    &   $0$\\
C$^+$ + O$_2$ $\rightarrow$ CO$^+$ + O   &    $3.42 \times 10^{-10}$    &    $0.00$    &   $0$\\
C$^+$ + O$_2$ $\rightarrow$ CO + O$^+$   &    $4.54 \times 10^{-10}$    &    $0.00$    &   $0$\\
C$^+$ + M $\rightarrow$ C + M$^+$   &    $1.10 \times 10^{-9}$    &    $0.00$    &   $0$\\
m$^+$ + M $\rightarrow$ m + M$^+$   &    $2.90 \times 10^{-9}$    &    $0.00$    &   $0$\\
H$^+$ + e$^-$ $\rightarrow$ H + $h\nu$   &    $3.50 \times 10^{-12}$    &    $-0.75$    &   $0$\\
He$^+$ + e$^-$ $\rightarrow$ He + $h\nu$   &    $5.36 \times 10^{-12}$    &    $-0.5$    &   $0$\\
H$_3^+$ + e$^-$ \begin{tabular}{l} $\rightarrow$ H + H + H \\  $\rightarrow$ H$_2$ + H\end{tabular}  &    $2.34 \times 10^{-8}$    &    $-0.52$    &   $0$\\
C$^+$ + e$^-$ $\rightarrow$ C + $h\nu$   &    $2.36 \times 10^{-12}$    &    $-0.29$    &   $0$\\
m$^+$ + e$^-$ $\rightarrow$ m$_1$ + m$_2$   &    $2.40 \times 10^{-7}$    &    $-0.69$    &   $0$\\
M$^+$ + e$^-$ $\rightarrow$ M + $h\nu$   &    $2.78 \times 10^{-12}$    &    $-0.68$    &   $0$\\
H$_2$ $\rightarrow$ H$_2^+$ + e$^-$ & $0.98\xi$ & &\\
H$_2$ $\rightarrow$ H$^+$ + H + e$^-$ & $0.02\xi$ & &\\
He $\rightarrow$ He$^+$ + e$^-$ & $0.53\xi$ & &\\
\hline
\end{tabular} 
\end{table}

\noindent {We mention that the two reactions H$_3^+$ + e$^-$ $\rightarrow$ H + H + H and H$_3^+$ + e$^-$ $\rightarrow$ H$_2$ + H occur at the same rate.}
\noindent The ionisation rate for potassium, sodium and hydrogen are given by \cite{PneumanMitchell} 
\begin{align}
\frac{dn_\mathrm{K^+}}{dt}= 6.5 \times 10^{-15} n_\mathrm{H_2} T^{\frac{1}{2}} \times e^{-\frac{5.1 \times 10^4\, \mathrm{K}}{T}} \mathrm{cm}^{-3}\, \mathrm{s}^{-1},\\
\frac{dn_\mathrm{Na^+}}{dt}= 1.4 \times 10^{-15} n_\mathrm{H_2} T^{\frac{1}{2}} \times e^{-\frac{6.0 \times 10^4\, \mathrm{K}}{T}} \mathrm{cm}^{-3}\, \mathrm{s}^{-1},\\
\frac{dn_\mathrm{H^+}}{dt}= 2.0 \times 10^{-10} n_\mathrm{H_2} T^{\frac{1}{2}} \times e^{-\frac{15.8 \times 10^4\, \mathrm{K}}{T}} \mathrm{cm}^{-3}\, \mathrm{s}^{-1}.
\end{align}

\noindent We considered recombination reactions at the surface of the grains. We used the collision rates of \cite{DraineSutin} and the interactions described in \citet{UmebayashiNakano1990}, \citet{IlgnerNelson} and \citet{KunzMouschovias2009}:
\begin{itemize}
\item Neutral grains: when hit by an electron, the electron sticks onto the grain with a probability of $P_e = 60\%$, while an ion always sticks.
\item Negatively charged grains: when hit by an ion, the ion recombines.
\item Positively charged grains: when hit by an electron, the grain becomes neutral.
\item Charged grains: if two grains with opposite charges collide, one electric charge is transferred.
\item Neutral grains: if hit by a charged grain, singly charged is transferred to the neutral grain.
\end{itemize}

The mean collision rate $\left< \sigma v\right>$ between a grain with a radius $a$ and a charge $q_\mathrm{g}$ and another species with a mass $m$ and a charge $q_\mathrm{s}$ is
\begin{equation}
\left< \sigma v \right> = \pi a^2  \left(\frac{8 k_\mathrm{B}T}{\pi m}\right)^{\frac{1}{2}}\left(1 - \frac{q_\mathrm{s}q_\mathrm{g}}{ak_\mathrm{B}T}\right)\left(1 + \left(\frac{2}{\frac{ak_\mathrm{B}T}{e^2}-2\frac{q_g}{q_s}}\right)^\frac{1}{2}\right)
\end{equation}
for $q_\mathrm{s}q_\mathrm{g} < 0$,
\begin{align}
\left< \sigma v \right> = \pi a^2  \left(\frac{8 k_\mathrm{B}T}{\pi m}\right)^{\frac{1}{2}}\left(1 + \left(\frac{4ak_\mathrm{B}T}{e^2} + 3\frac{q_\mathrm{g}}{q_\mathrm{s}}\right)^{-\frac{1}{2}}\right)^2 \nonumber \\
        \times  \exp \left(- \frac{q_\mathrm{g}e^2}{(q_\mathrm{s}ak_\mathrm{b}T)\left(1+\left(\frac{q_\mathrm{s}}{q_\mathrm{g}}\right)^{\frac{1}{2}}\right)}\right)
\end{align}
for $q_\mathrm{s}q_\mathrm{g} > 0$ and
\begin{equation}
\left< \sigma v \right> = \pi a^2  \left(\frac{8 k_\mathrm{B}T}{\pi m}\right)^{\frac{1}{2}}\left(1 + \left(\frac{\pi e^2}{2ak_\mathrm{B}T}\right)^{\frac{1}{2}}\right).
\end{equation}
for $q_\mathrm{g} = 0$.

For two grains of opposite charges $q$ and $q^{'}$ with radii $a$ and $a^{'}$ and a reduced mass $\mu_\mathrm{g}=\frac{m m^{'}}{m+m^{'}}$, the collision rate is
\begin{equation}
\left< \sigma v \right> = \pi (a + a^{'})^2  \left(\frac{8 k_\mathrm{B}T}{\pi \mu_\mathrm{g}}\right)^{\frac{1}{2}}\left(1 - \frac{qq^{'}}{(a+a^{'})k_\mathrm{B}T}\right),
\end{equation}
and
\begin{equation}
\left< \sigma v \right> = \pi (a + a^{'})^2  \left(\frac{8 k_\mathrm{B}T}{\pi \mu_\mathrm{g}}\right)^{\frac{1}{2}}\left(1 + \left(\frac{\pi e^2}{2(a+a')k_\mathrm{B}T}\right)^{\frac{1}{2}}\right)P_a.
\end{equation}
between a charged and a neutral grain. $P_a = \frac{a^2}{a^2+a^{'^2}}$ denotes the probability of a charge transfer to the neutral grain of radius $a^{'}$.

The abundances of the neutrals are also taken from \citet{UmebayashiNakano1990} (\citet{KunzMouschovias2009} for potassium). Because of its high thermal ionisation rate, K is the only neutral species whose fractional abundance varies in our network.

\begin{table}
\label{quantities}
\caption{Abundances of neutrals relative to H.}
\centering
\begin{tabular}{lr}
\hline\hline
Species & Relative abundance\\
\hline
H$_2$ & $0.5$\\
He & $8.5 \times 10^{-2}$\\
C & $8.4 \times 10^{-5}$\\
O & $4.1 \times 10^{-5}$\\
O$_2$ & $4.8 \times 10^{-5}$\\
M & $1.7 \times 10^{-6}$\\
K & $2.2 \times 10^{-10}$\\
Na & $3.1 \times 10^{-9}$\\
\hline
\end{tabular}
\end{table}

\section{Grain charges \label{Charge_appendix}}

\begin{figure}
\begin{center}
\includegraphics[width=0.5\textwidth]{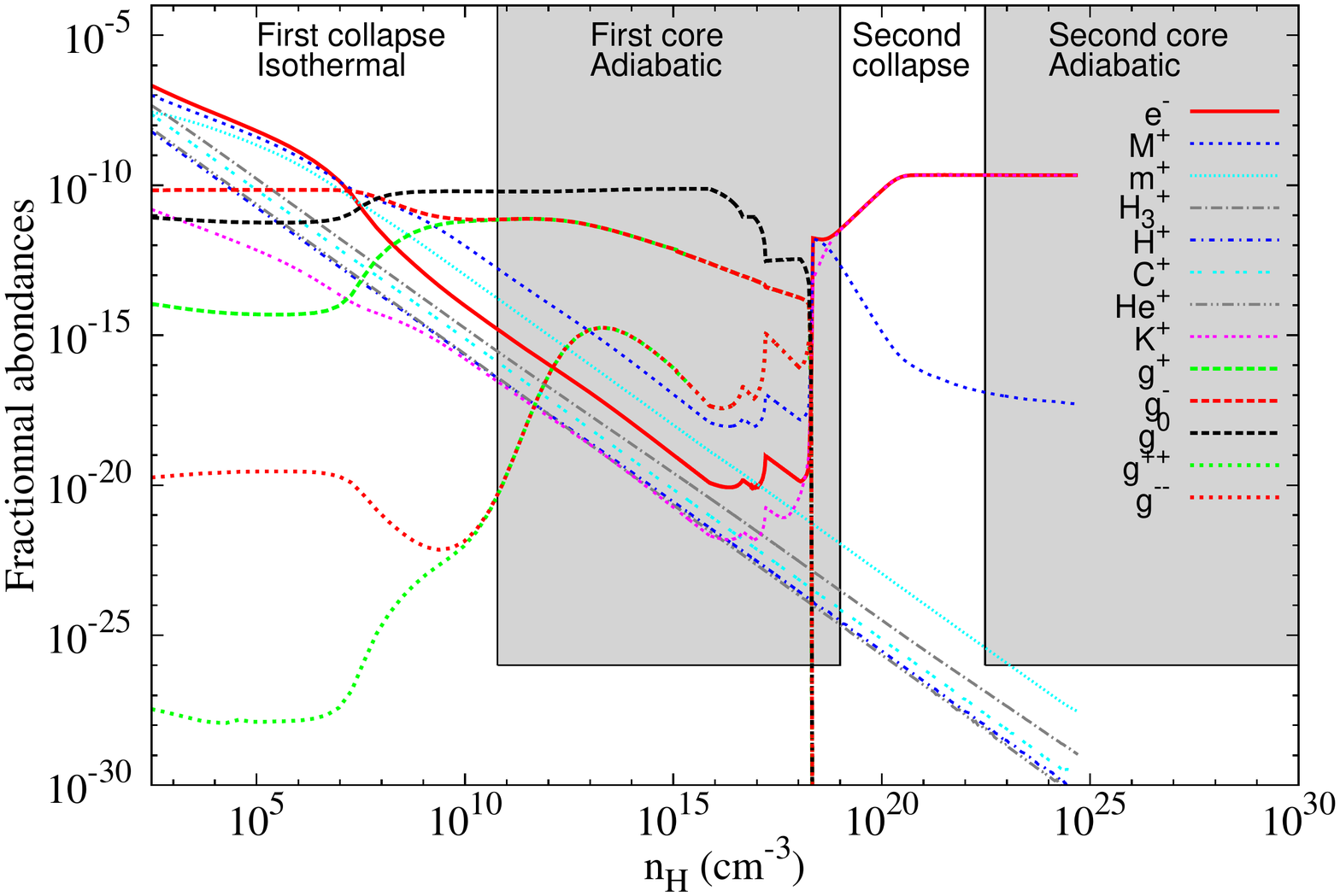}
\caption{Abundances of the chemical species with double-charge grains, without Na, the thermal ionisation of H and the thermionic emission.}
\label{2charges}
\end{center}
\end{figure}

\begin{figure}
\begin{center}
\includegraphics[width=0.55\textwidth]{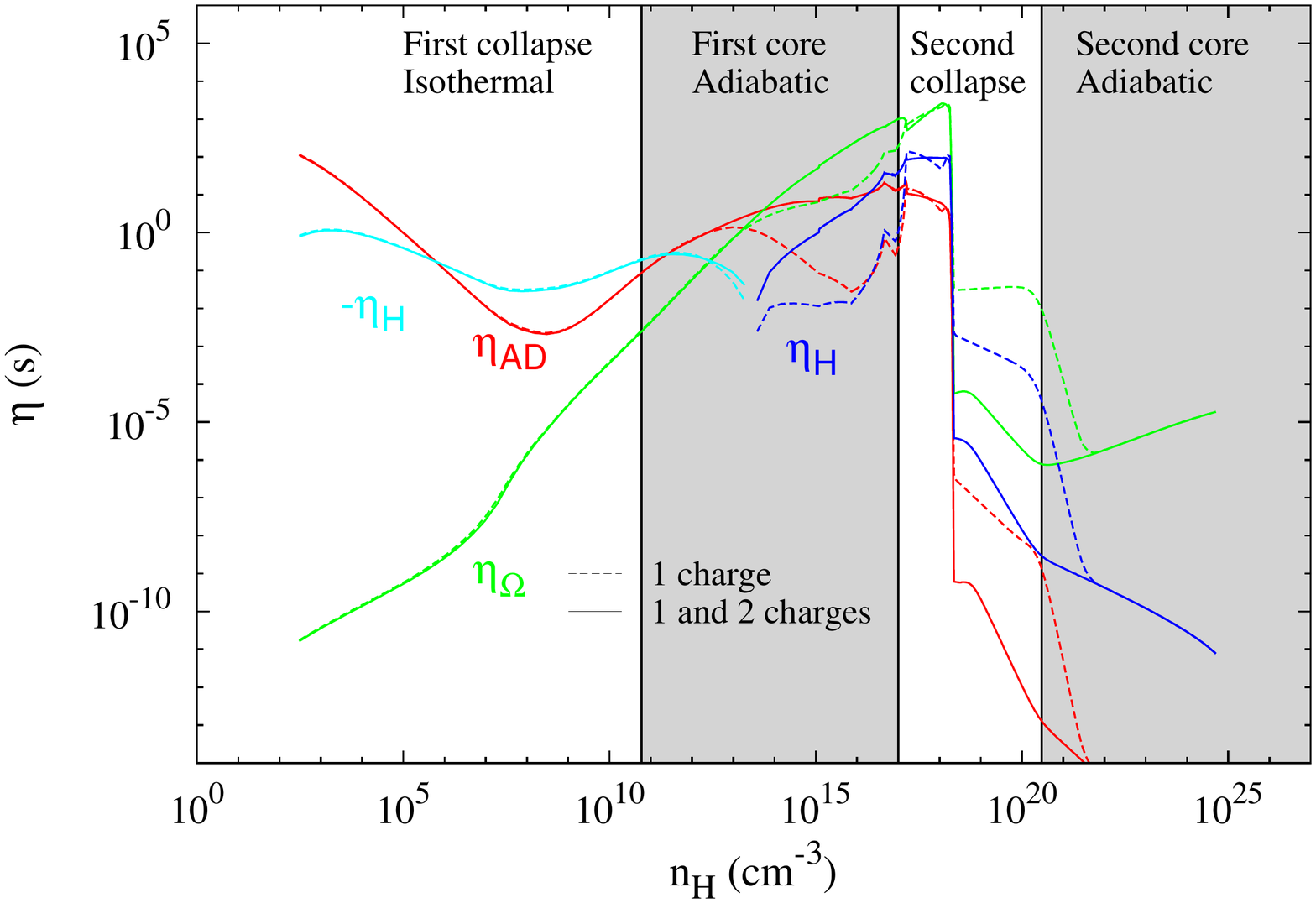}
\caption{Resistivity comparison for the simplified case. Dashed line: network with only singly charged grains, solid line: network with singly and doubly charged grains.}
\label{comp_eta_charges}
\end{center}
\end{figure}

Multiply charged grains are part of a chemical network \citep{NishiNakanoUmebayashi1991}. Figure \ref{2charges} shows the abundances of species in a simplified case (only one bin of size, without Na, the thermionic emission and thermal ionisation of H) but with grains holding two electric charges. Although they are generally less abundant than single-charge grains, they effect the charge distribution, especially for $n_\mathrm{H} > 10^{12}$cm$^{-3}$. The main effect is that one-charge grains, the main charge carriers, seem to form more neutral grains and are less numerous than in the fiducial case. This of course affects the resistivities, as shown in \refig{comp_eta_charges}.

\begin{figure}
\begin{center}
\includegraphics[trim=2cm 0cm 0cm 0cm, width=0.55\textwidth]{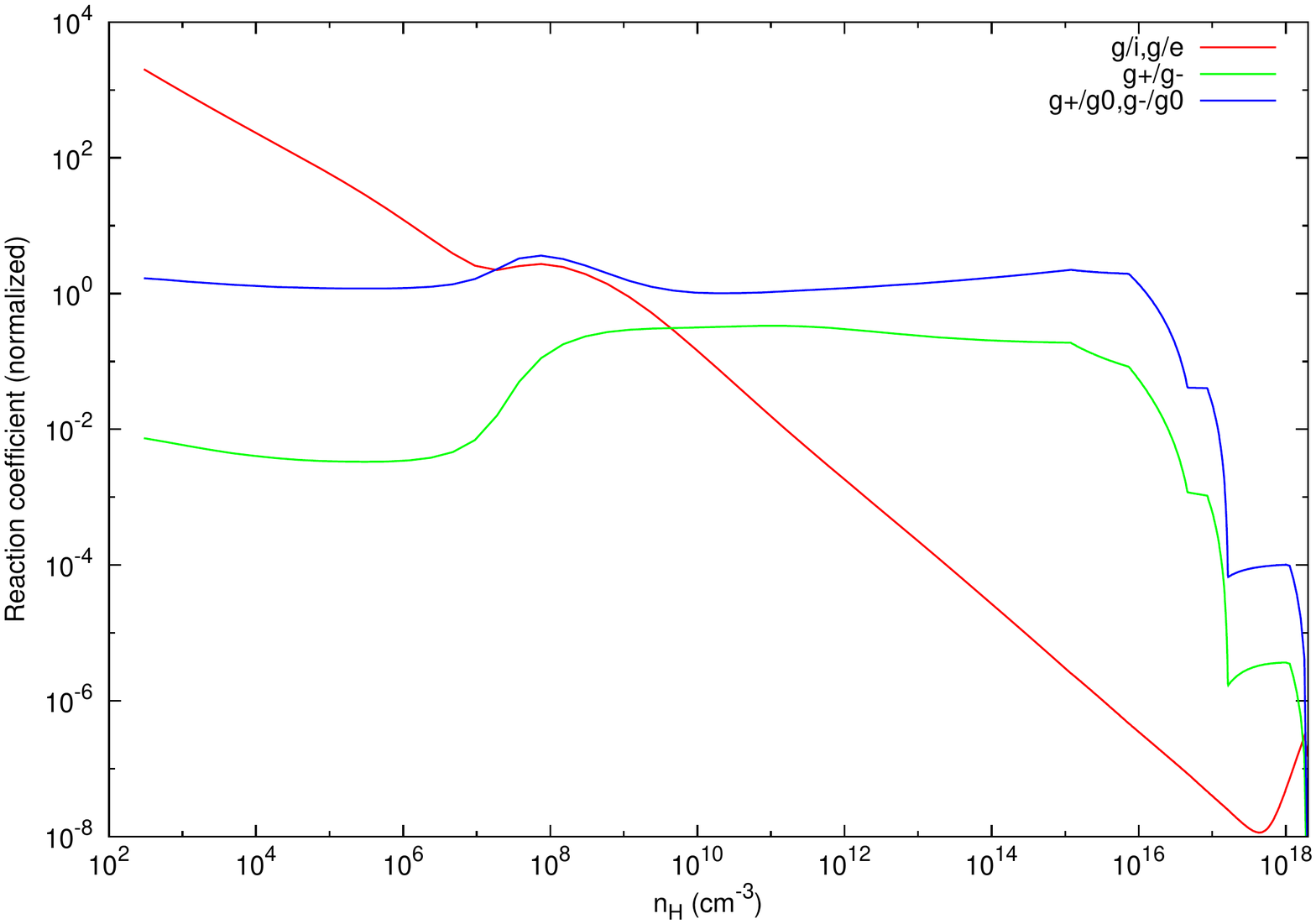}
\caption{Highest reactions rates for grain/ion and grain/electron reactions (red curve), positively charged grain/negatively charged grain reactions (green curve) and neutral grain/charged grain reactions (blue curve).}
\label{Coefs}
\end{center}
\end{figure}

\begin{figure}
\begin{center}
\includegraphics[width=0.45\textwidth]{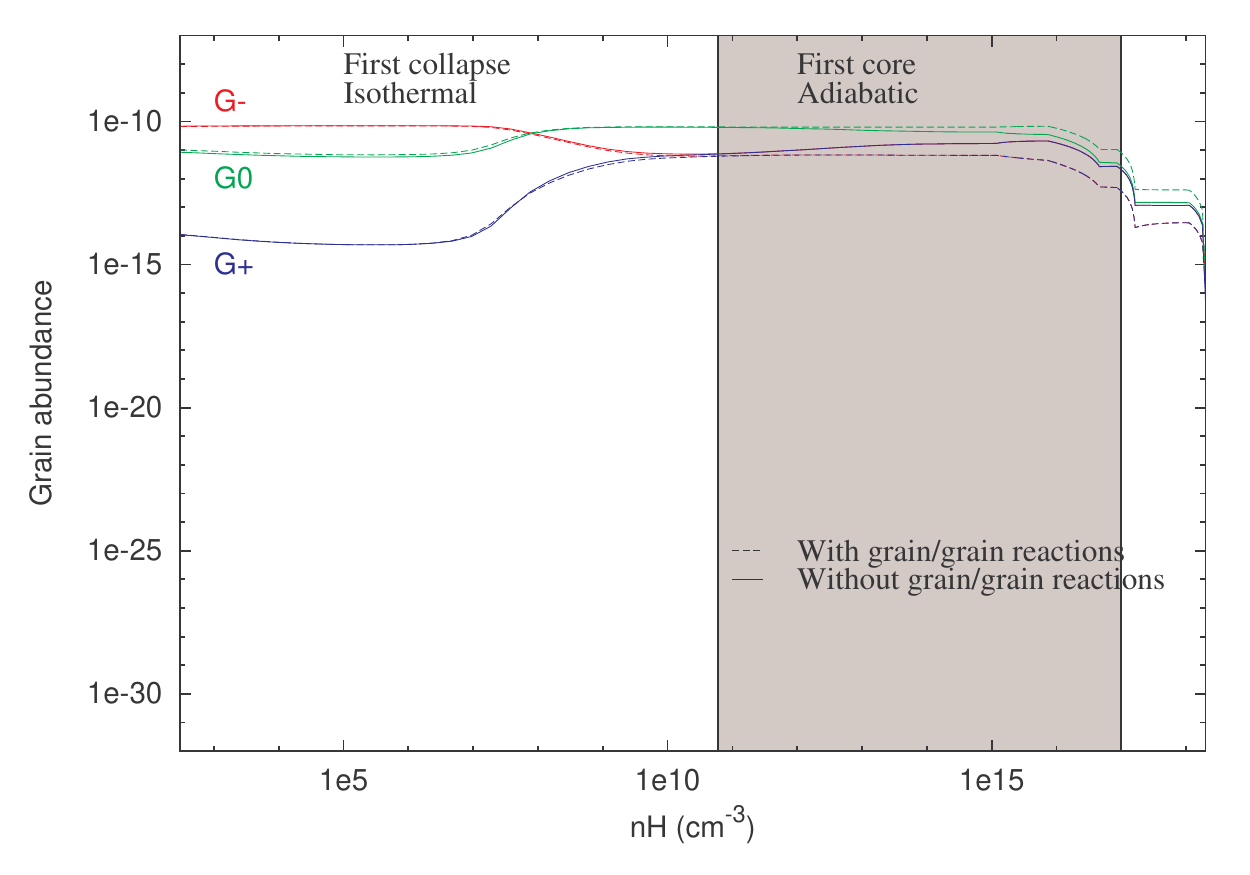}
\includegraphics[width=0.45\textwidth]{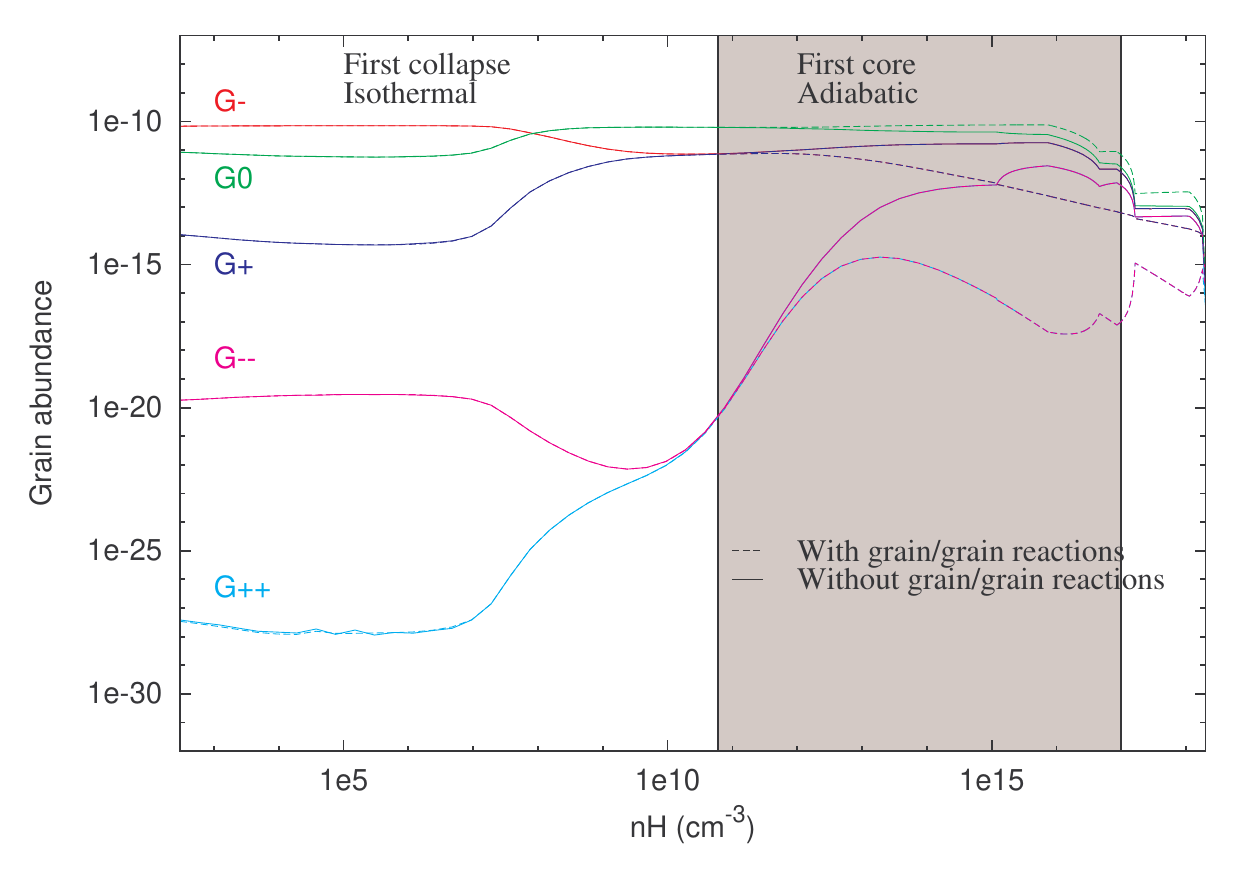}
\caption{Comparison of the grain abundances with (dashed lines) and without (solid lines) charge transfer between grains for singly charged grains (top figure) and singly and doubly charged grains (bottom figure).}
\label{grain_transfornot}
\end{center}
\end{figure}

We were only able to produce such a result for one bin of size with the current method of resolution. The abundance gap between the two-charge holding grains and the remaining the species spans over 20 orders of magnitude (especially at low densities in this case). This gap widens for several number of bins because small grains tend to hold fewer charges, and large grains are less abundant. Dealing with so many orders of magnitude is problematic because numeric round-off errors may artificially change the contribution of the less abundant species. Several authors have used analytical expressions of charge distribution to avoid this numerical difficulty \citep{DraineSutin,Okuzumi2009,Fujii2011}. These models, however, do not take the charge transfer between grains into account. \refig{Coefs} shows the highest reaction rates of grain-ion, grain-electron and grain-grain reactions. Reaction rates between grains are largely dominant for $n_\mathrm{H} > 10^{12}$cm$^{-3}$, and not taking them into account leads to large discrepancies in the grains abundances in this density range (see \refig{grain_transfornot}), with in turn affects the resistivities.

\begin{figure}
\begin{center}
\includegraphics[width=0.5\textwidth]{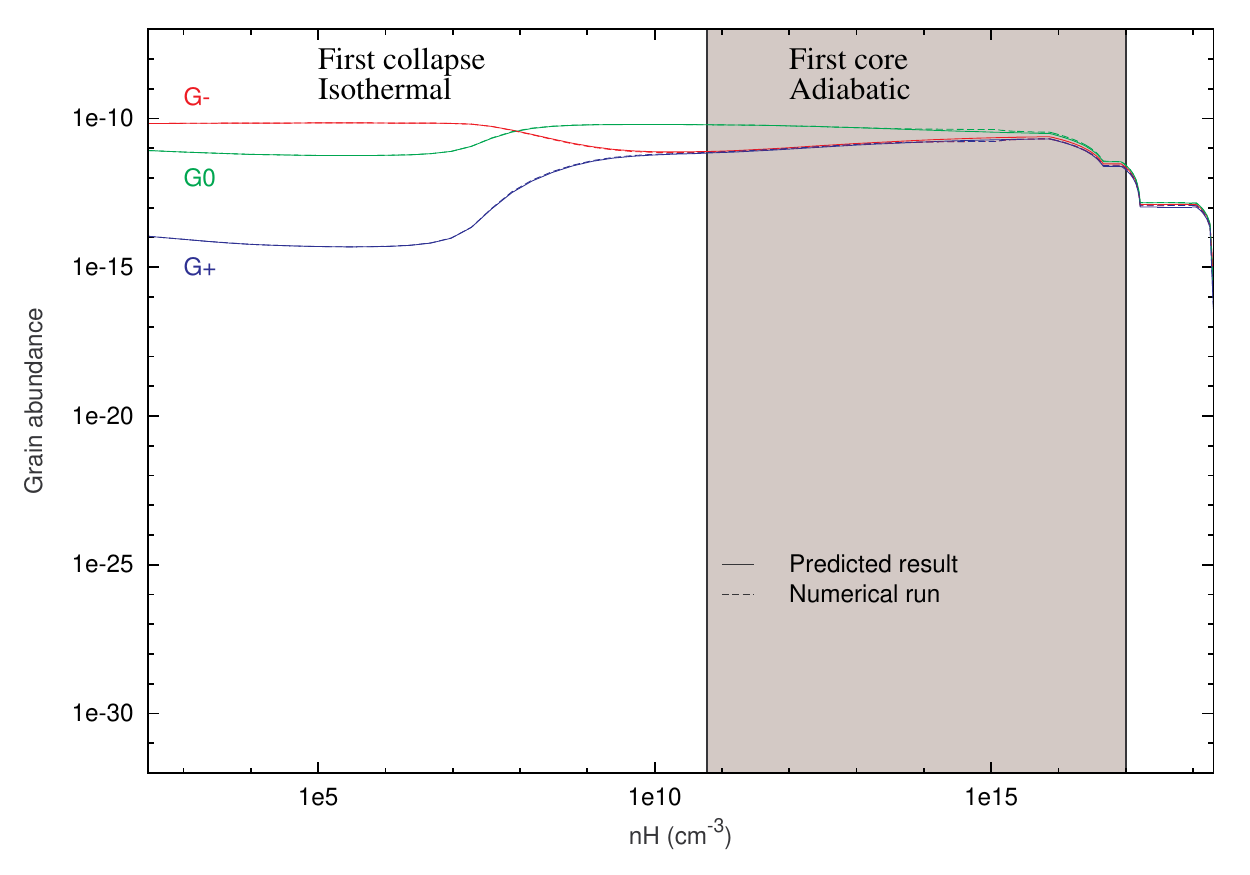}
\includegraphics[width=0.5\textwidth]{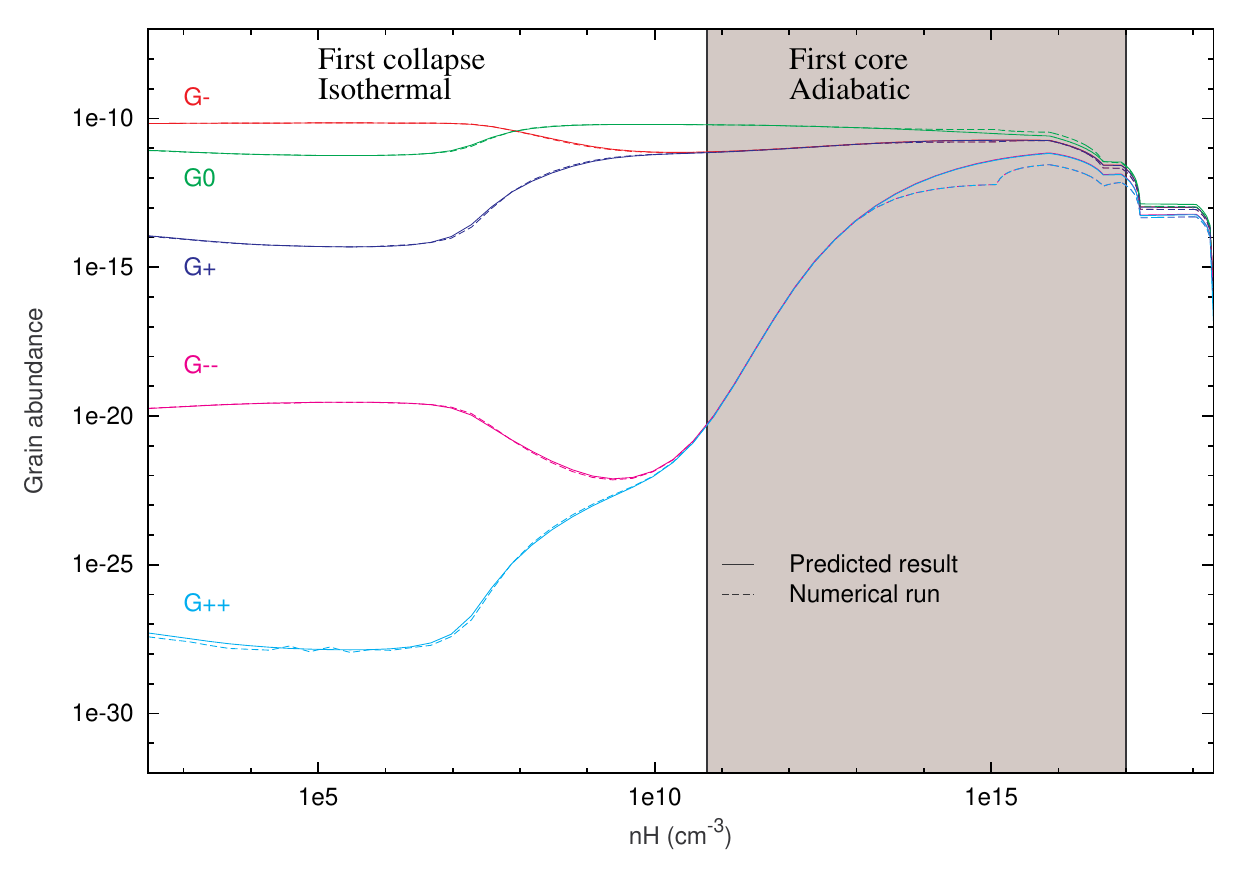}
\caption{Comparison of the abundances of grains computed with our code (dashed lines) and predicted from the ions and electron abundances (solid lines) for singly charged grains (top figure) and singly and doubly charged grains with one bin (bottom figure).}
\label{predict}
\end{center}
\end{figure}

\begin{figure}
\begin{center}
\includegraphics[width=0.5\textwidth]{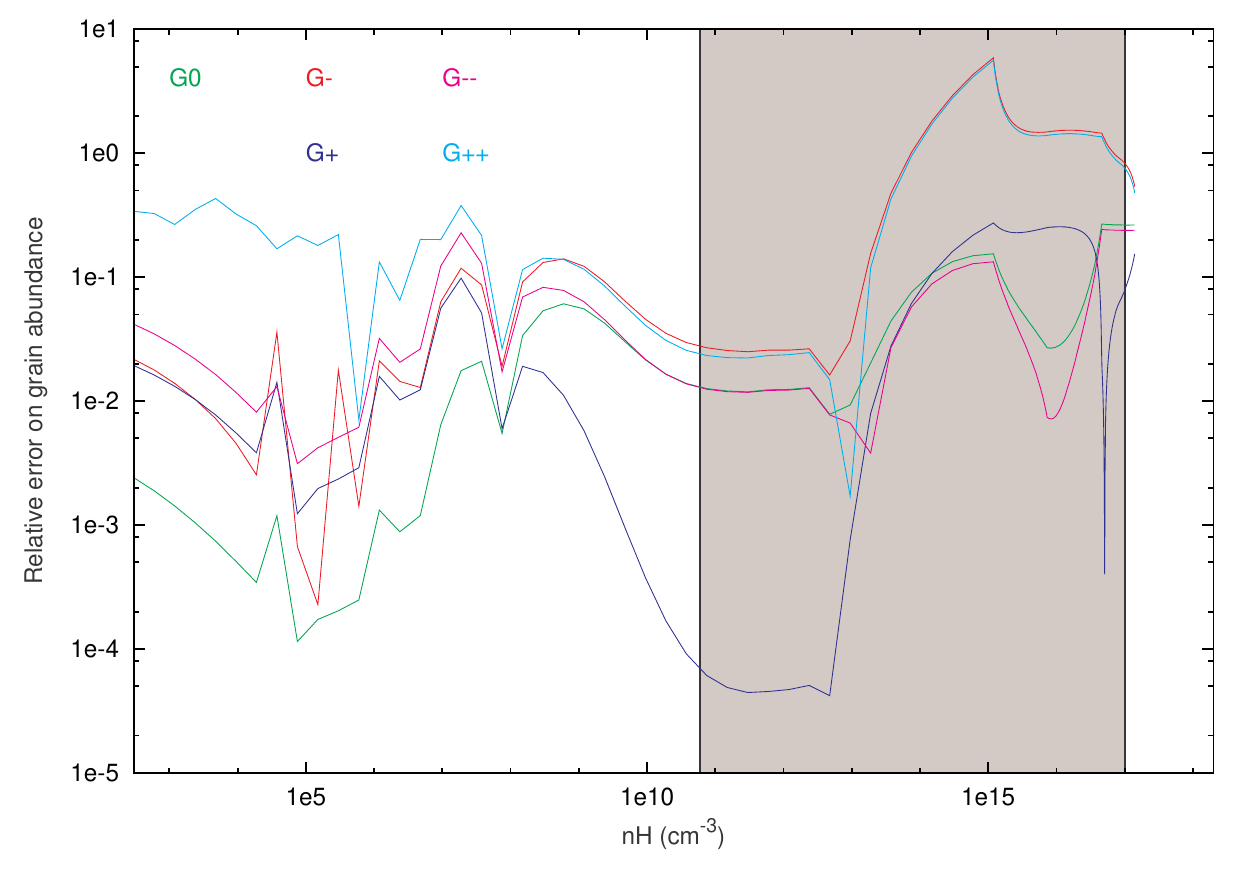}
\caption{Relative error $\frac{\mathbf{x}_\mathrm{predict} - \mathbf{x}_\mathrm{num}}{\mathbf{x}_\mathrm{num}}$ between the numerical abundances of grains and the predicted results of \refig{predict} for singly and doubly charged grains.}
\label{error_predict}
\end{center}
\end{figure}

\begin{figure}
\begin{center}
\includegraphics[width=0.5\textwidth]{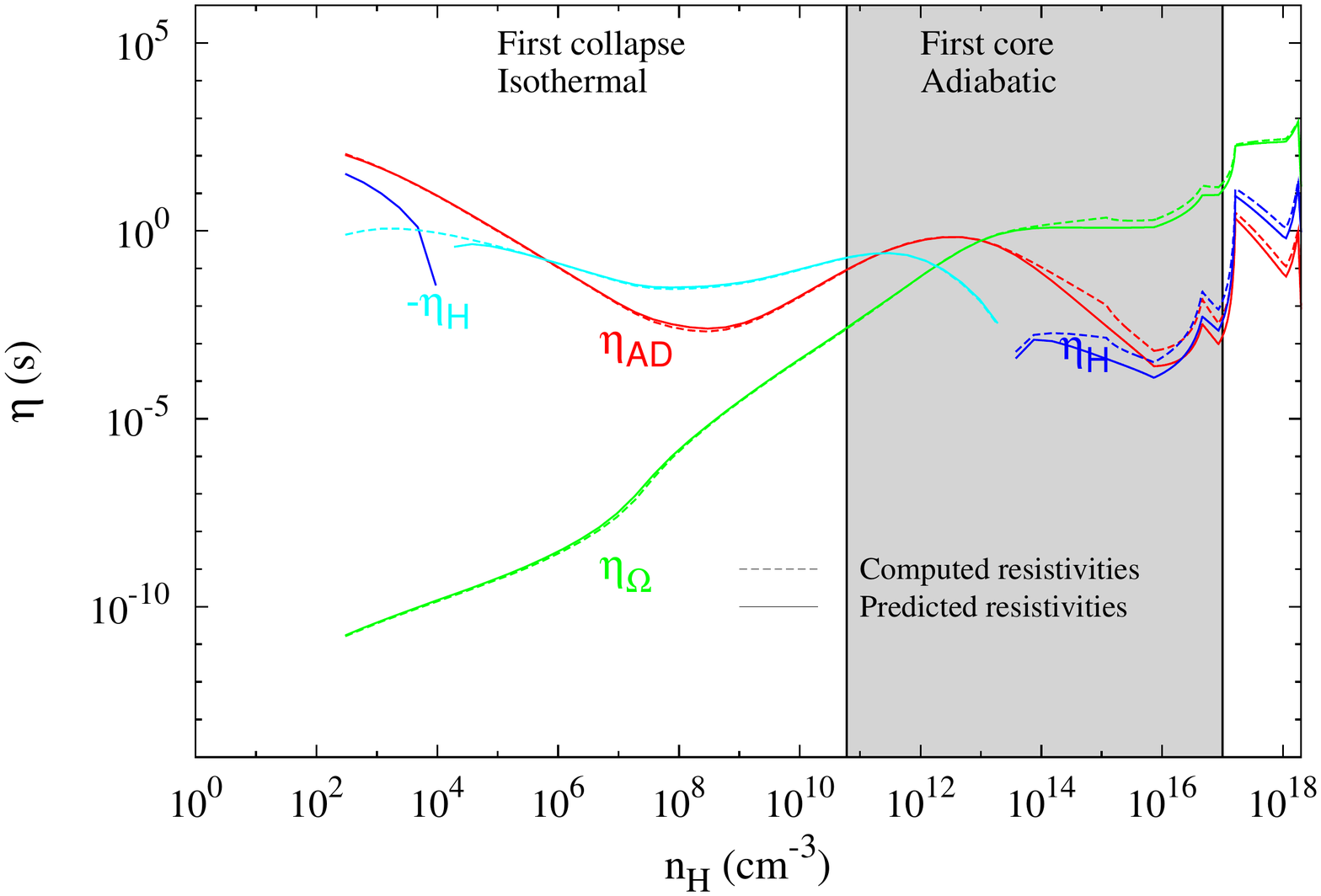}
\caption{Comparison of the resistivities using abundances computed with our code (dashed lines) and abundances predicted with the \citet{DraineSutin} formulae (solid lines) for singly and doubly charged grains, without charge transfer between grains.}
\label{comp_2chg_predict_eta}
\end{center}
\end{figure}

On the other hand, using the analytical expressions of \citet{DraineSutin} with the ion and electron abundances calculated in our simplified case with only one charge and without considering grain-grain reactions enables us to predict the grain abundances for one and two charges with good precision. \refig{predict} depicts these predicted abundances of grains, compared to those computed with our code, for one and two charges. The abundances of the doubly charged grains are slightly overestimated for $10^{14} < n_\mathrm{H} < 10^{17}$cm$^{-3}$, but the agreement is very good otherwise. \refig{error_predict} shows the relative error  $\frac{\mathbf{x}_\mathrm{predict} - \mathbf{x}_\mathrm{num}}{\mathbf{x}_\mathrm{num}}$ of the model prediction for the numerical run with two charges allowed. All abundances agree within roughly 10\%, except for g$^{++}$ at low densities and g$^-$ and g$^{++}$ for $n_\mathrm{H} > 10^{12}$cm$^{-3}$. \refig{comp_2chg_predict_eta} shows the corresponding resistivities. Both models are relatively similar except in the density ranges mentioned above. For $10^{14} < n_\mathrm{H} < 10^{17}$cm$^{-3}$, the model resistivities are slightly underestimated, but the differences with the numerical results remain similar. At low densities, the errors on the abundances result in a sign change of the Hall resistivity because even a small change allows the positive contribution of $\eta_\mathrm{H}$ to overcome the negative contribution. However, the Hall effect is not expected to play a role at these early stages of protostellar collapse. Therefore, even though we were unable to verify the exactness of the prediction for grains holding a larger number of charges (because of numerical difficulties), it seems reasonable to say that our method gives satisfying results at least for a small number of charges, which is the relevant case because grains holding more than three charges are not expected to be abundant enough to significantly modify the resistivities \citep{Sano2004}.

\section{Details of the numerical method}

The Jacobian general expression (for charged particles alone, neutral grains are taken care of separately) reads:
\begin{align}
\mathbb{J}_{ij} &= \frac{\partial F_i}{\partial x_j} \nonumber \\
 &=\alpha_{ij} + \sum_{k=1}^N \beta_{ijk} x_k - \gamma_{ij} x_i -\delta_{ij} \sum_{k=1}^N \gamma_{ik} x_k.
\end{align}
The Jacobian expression for neutral grains in the bin $\alpha$ is:
\begin{align}
\mathbb{J}_{g_0^{\alpha}j} = -\mathbb{J}_{g_+^{\alpha}j}-\mathbb{J}_{g_-^{\alpha}j}.
\end{align}
This makes the Jacobian matrix $\mathbb{J}$ singular, which is a great problem when solving the system using Newton-Raphson iterations. For the semi-implicit method, however, a simple solution consists of reducing the calculated species to charged particles alone and then updating with the neutral grains at the end of the iteration.\\

\begin{figure*}
\begin{center}
\includegraphics[width=0.85\textwidth]{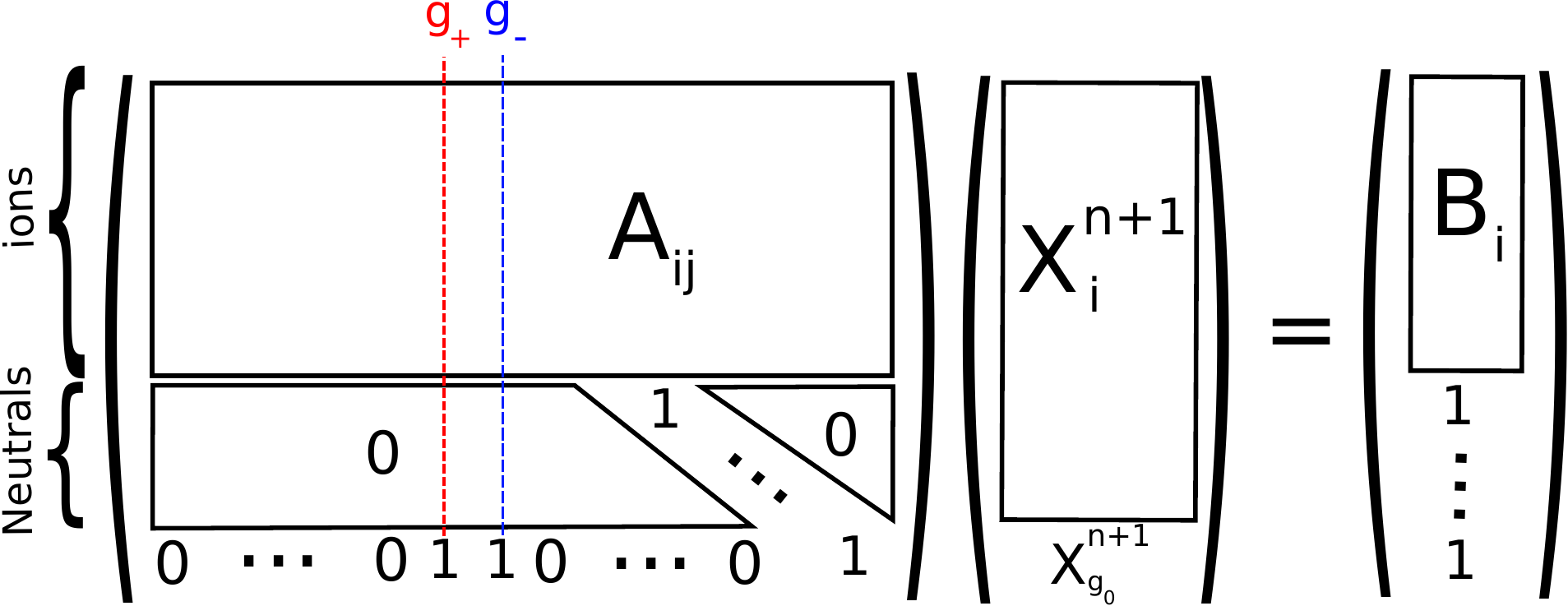}
\caption{Matrix visualisation of the system to be solved after linearisation in time.}
\label{matrix}
\end{center}
\end{figure*}

\noindent A schematic diagram of the system is shown in \refig{matrix}, with
\begin{align}
A_{ij} &= \delta_{ij} - \tilde{dt} \left[ \alpha_{ij} - \frac{n_\mathrm{H}}{\zeta} \gamma_{ij} x_i^n + \frac{n_\mathrm{H}}{\zeta} \sum_{k=1}^N (\beta_{ijk} - \gamma_{ik} \delta_{ij}) x_k^n \right],  \\
B_{i} &= \frac{n_\mathrm{H}}{\zeta} \left[ \sum_{j=1}^N \gamma_{ij} x_j^n x_i^n - \frac{1}{2} \sum_{j=1}^N \sum_{k=1}^N \beta_{ijk} x_j^n x_k^n  \right]\tilde{dt}  + x_i^n.
\end{align}

 While the system nearly reaches equilibrium, the abundances of the least abundant species may continue to vary significantly in relative values 
for a long time. For this reason, we chose to stop the calculations when the time-step $\tilde{dt}$ reached a final constant value. 
The least abundant species may continue to evolve, but this is inconsequential for evaluating the equilibrium abundances of the dominant species.

\section{Code validation}

\begin{figure}
\begin{center}
\includegraphics[trim= 2cm 2cm 2cm 2cm, width=0.5\textwidth]{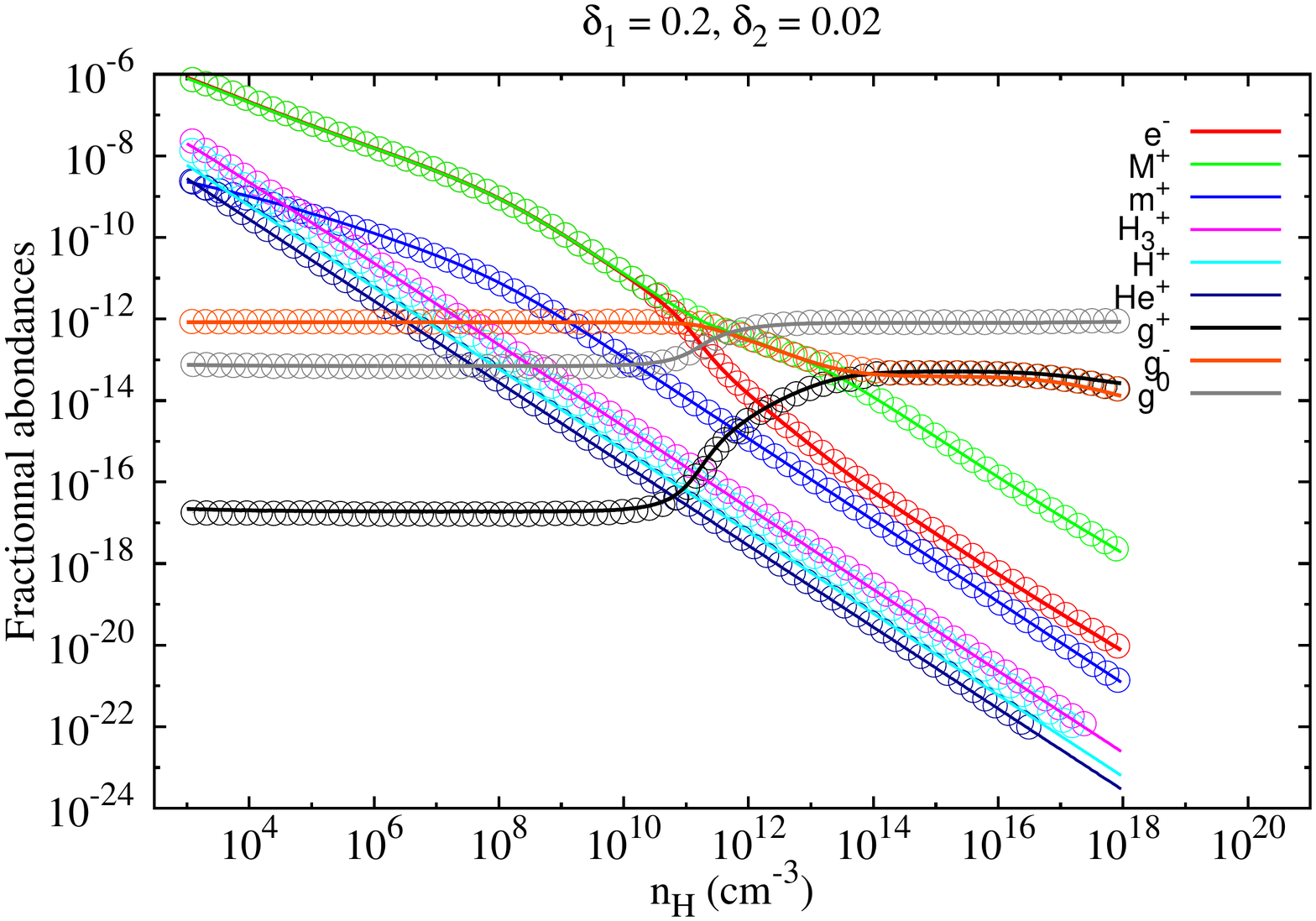}
\includegraphics[trim= 2cm 2cm 2cm 2cm, width=0.5\textwidth]{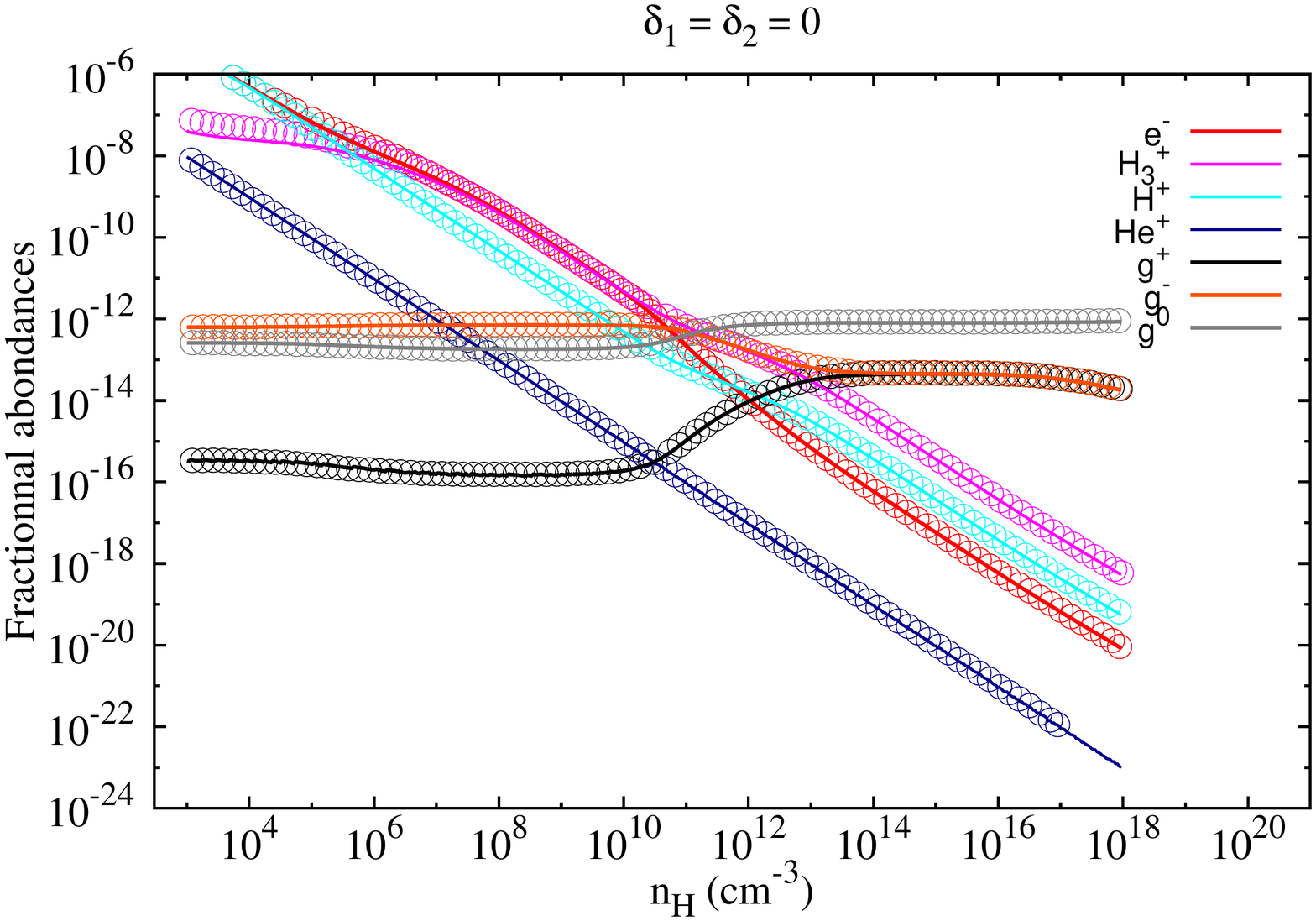}
\caption{Evolution of the abundances of various charged particles as a function of the number density of hydrogen atoms, at $T=10$ K. Solid lines are our results, and circles are taken from Figs. 2 and 4 of \citet{UmebayashiNakano1990}. Top: $\delta_1 = 0.2$, $\delta_2 = 0.02$. Bottom: $\delta_1 = \delta_2= 0$.}
\label{UN90}
\end{center}
\end{figure}

In order to test the code with the most simple parameter set, we compared it with the results of \citet{UmebayashiNakano1990}, which did not include potassium and grain evaporation. We considered the basic case of one bin of grains, with a size a$_0$. Let $\delta_1$ be the fraction of C and O in the gas phase, and $\delta_2$ the fraction of metals. Our calculations are shown in \refig{UN90}, with $\delta_1=0.2$ and $\delta_2=0.02$ for the top panel, and $\delta_1=\delta_2=0$ for the bottom panel. It represents the evolution of the relative abundances with the density at $T=10$ K, compared with data from \citet{UmebayashiNakano1990}. It shows that the two agree very well. For $n_\mathrm{H} < 10^{10}$ cm$^{-3}$, most of the grains have a negative charge and electrons and ions are the dominant species. For $n_\mathrm{H} > 10^{13}$ cm$^{-3}$, charged grains are more abundant than electrons and ions.

\end{appendix}

\bibliographystyle{aa}
\bibliography{MaBiblio}

\end{document}